\documentclass[showpacs,prc]{revtex4}
\usepackage{dcolumn}
\usepackage{amsmath}
\makeatletter
\providecommand{\LyX}{L\kern-.1667em\lower.25em\hbox{Y}\kern-.125emX\@}

\usepackage[T1]{fontenc}
\usepackage[latin1]{inputenc}
\usepackage{graphics}
\usepackage{comment}
\makeatother

\begin{document}

\newcommand{\be}{\begin{eqnarray}}
\newcommand{\ee}{\end{eqnarray}}
\topmargin -1cm

\title{ Few body impulse and fixed scatterer
approximations for high energy scattering}

\author{R. Crespo}
\email{raquel.crespo@tagus.ist.utl.pt}
\affiliation{Departamento de F\'{\i}sica,
Instituto Superior T\'ecnico,\\
Av. Prof. Cavaco e Silva,  Taguspark,
2780-990 Porto Salvo, Oeiras, Portugal}
\affiliation{Centro de F\'{\i}sica Nuclear, Av Prof. Gama Pinto, 2,
1699, Portugal}
\author{A.M. Moro}
\email{moro@us.es}
\affiliation{Departamento de F\'{\i}sica,
Instituto Superior T\'ecnico,\\
Av. Prof. Cavaco e Silva,  Taguspark,
2780-990 Porto Salvo, Oeiras, Portugal}
\affiliation{Departamento
de F\'{\i}sica At\'omica, Molecular y Nuclear,
Universidad de Sevilla,
Apdo. 1065, E-41080 Sevilla, Spain}
\author{I.J. Thompson}
\affiliation{Physics Department University of Surrey\\
Guildford, Surrey, GU2 7XH, U.K.}

\date{\today}
\begin{abstract}
The elastic scattering  differential cross section is calculated
for proton scattering from $^6$He at 717 MeV, using single
scattering terms of the multiple scattering expansion of the total
transition amplitude (MST).
We analyse the effects of different
scattering frameworks, specifically the Factorized Impulse
Approximation (FIA)  and the Fixed Scatterer (adiabatic)
Approximation (FSA) and
the uncertainties associated with the use
different structure models.

\end{abstract}

\maketitle

\section{Introduction}

The understanding of nuclei can only be fully achieved by studying
the way they interact with other nuclei. This is particularly true
for halo nuclei, due to their short lifetime and  simplified energy
spectrum. We consider here the scattering of a structureless
projectile by a target assumed to be composed of structureless
subsystems (as is approximately the case for halo nuclei).

It is the aim of a microscopic reaction theory to construct the
total scattering amplitude in terms of well defined dynamical and
structural quantities.
For the scattering of a nucleon by a target with ${\cal N}$ particles, one
has to solve a $n= {\cal N}+1$ many body scattering problem.
For ${\cal N}=2$ clusters of equal mass this has been done solving
the Faddeev equations \cite{Joa87}
and generalized  for  ${\cal N}=3$  \cite{Alt70,Fon84}.
This  many body scattering framework is however very complicated,
and does not handle clusters of different mass,
and so alternative methods have  been developed as for example
the Continuum  Discretized Coupled Channels (CDCC).
In this approach  a system of coupled equations needs to be solved
with effective projectile-subsystem interactions. This method  has been
successfully applied for more than two decades to the scattering of
(${\cal N}=2$)-body  targets  for a wide range of projectile  masses
and bombarding energies.
The case of the scattering from a (${\cal N}=3$)-body system
is considerably more demanding, although some work is already
in progress \cite{Mat04}.
Alternatively in the high energy regime
one can use a multiple scattering expansion of the
total transition amplitude, MST \cite{Gol64,Wat57,Joa87,Cres99,Cres02,CresRNB}.

When describing the scattering of stable from halo nuclei,
it is crucial to model the halo many-body
character of ${\cal N}$ composite particles \cite{Cres99,Cres02,CresRNB}.
In particular, a three or four-body problem has to be solved when
studying the scattering of a projectile by a target which is
a bound state of two or three subsystems, as is the
case for $^{11}$Be and $^6$He respectively.

This problem can be conveniently addressed  by the MST approach.
In this formalism, the projectile-target transition amplitude
is expanded in terms of off-shell transition amplitudes for
projectile-subsystem scattering.
Due to the complexity of the many-body operator, suitable
approximations  need to be made in
order to express in a convenient way
the overall scattering amplitude in terms of the
scattering by each target subsystem. Under further suitable
approximations each term
can be written in terms of a product of a form factor and the
transition amplitude for the scattering for that subsystem.
The MST  method provides  a clear and transparent
interpretation of the scattering of a composite system in terms of the
free scattering of its constituents and is numerically advantageous.
This scattering framework
is particularly useful at high energies and for loosely bound nuclei
where  the expansion is expected to converge quickly.

In this paper, we use this framework to calculate the elastic scattering
differential cross section  for proton scattering on $^6$He at 717 MeV where
new data has been obtained at higher transferred momentum than previously.
These new momentum transfers, however, still essentially probe only that part
of the few-body dynamics of the halo cluster which is constrained by the rms
radius. Therefore the few-body treatment of the halo is appropriate in this
energy and angular range, and we will see that calculated differential cross
sections for various structure models largely reflects properties
closely connected to rms radii. It is the aim of the present paper to
clarify and test the approximations involved in the applying the MST multiple
scattering expansion to find the predicted cross sections for proton-halo
elastic scattering at these momentum transfers. Specifically, we examine (i)
the impulse approximation, (ii) the factorization approximations and  (iii) the
single scattering approximation.

\section{Multiple scattering expansion of the total transition amplitude}

We consider the scattering of a projectile
(system 1) from a few-body target consisting of ${\cal N}$
sub-systems weakly bound to each other.
We shall frequently refer to the composite system as the
target although in practice an actual experiment may be carried out with the
 composite system   as a projectile.
The subsystems ${\cal I}, {\cal J}, \ldots$ are assumed to be stable
and can be either composite nuclei or nucleons.
The total transition  amplitude for the scattering is
\be
T &=& V + V G T \nonumber \\
& = &  \sum_{{\cal I}=2}^{n} v_{\cal I} + \sum_{{\cal I}=2}^{n} v_{\cal I}
G T~~,
 \label{Tsum}
\ee
with $n = {\cal N} + 1$, $v_{\cal I}$  the interaction between the projectile
and the ${\cal I}$ target subsystem, and $G$ is the  propagator
\be
G = \left( E^+ - K - \sum V_{{\cal I}{\cal J}} \right)^{-1}.
\label{propagatorG}
\ee
Here, $E$ is the total energy and is related to the total incident kinetic
energy
$E_1 = \frac{\hbar^2 k_1^2}{2 \mu_{NA}}$ by $E=E_1+\epsilon_0$ where
$\epsilon_0$
is the target ground state energy. At this stage we use non-relativistic
kinematics. The inclusion of relativistic kinematics will be discussed later.
 The operator
$K$ corresponds to  the total kinetic energy of the projectile and ${\cal N}$
target sub-systems
in the projectile-target center of mass frame.
In Eq.~(\ref{propagatorG}) $V_{{\cal I}{\cal J}}$ is the interaction
between   subsystems ${\cal I}$ and ${\cal J}$.
Equivalently, we may write the propagator in terms of the kinetic
energy operator for the projectile $K_1$ in the center of mass
of the interacting projectile-target (P-T) system, and the target
nucleus Hamiltonian, $H_0$
\be
G = \left( E^+ - K_1 - H_0  \right)^{-1}.
\label{propagatorGopt}
\ee
The total transition amplitude,  Eq.(\ref{Tsum}), can be rewritten as
\be
T =  \sum_{\cal I}   \tau_{\cal I} +  \sum_{\cal I}    \tau_{\cal I} G
 \sum_{{\cal J} \neq {\cal I}}  \tau_{\cal J} + \cdots ~~.\label{Tfullexp}
\ee
where the projectile-${\cal I}$ subsystem transition amplitude
$\tau_{\cal I}$ is given by
\be
\tau_{\cal I} = v_{\cal I} + v_{\cal I} G \tau_{\cal I}~~.\label{tau}
\ee
We note that the  propagator in $\tau_{\cal I}$ contains the
target Hamiltonian and thus it is still at this stage a many-body operator.
In the limit when the target nucleus
subsystems are weakly bound to each other, the multiple scattering expansion
to the P-T transition amplitude  is expected to converge
rapidly \cite{Wat57,Gol64} and the MST  expansion
Eq.~(\ref{Tfullexp}) can be used.

We apply next this formalism to proton scattering from a target
of two and three bound structureless subsystems.

\section{The Single scattering 3-body problem}

We first consider the  scattering of a projectile of mass $m_1$
from a target (such as $^{11}$Be) assumed to be well described by a
two body model with two subsystems (of valence particle and a core)
labeled here as 2 and 3 of masses $m_2$ and $m_3$ respectively.
Let $\vec{k}_1, \vec{k}_2, \vec{k}_3$, ($\vec{k}'_1, \vec{k}'_2, \vec{k}'_3$)
be the initial (final)  momenta of the
projectile and the two cluster subsystems.

Neglecting core excitation, the  target wave function  is
\be
\Phi(\vec{r},{\xi}_3)
= \left[ \phi_{23}(\vec{r}) \otimes
 \varphi_{3}({\xi}_3) \right] ~~, \label{FBwf1}
\ee
where $\varphi_{3}({\xi}_3)$ is the core internal wave function
and  $\phi_{23}$ is the wave function describing the relative motion of the
(2,3) pair.

The elastic transition amplitude to first order in the projectile-subsystem
transition amplitudes is
\be
T &=& \langle \vec{k}'_1 \Phi | \tau_{2} | \vec{k}_1 \Phi \rangle
+ \langle \vec{k}'_1 \Phi | \tau_{3} | \vec{k}_1 \Phi \rangle
+ \cdots
\label{t2bodyexp}
\ee

There are two approaches in handling the dynamics of the few-body
system: The impulse (I) and the Fixed Scatterer (or frozen halo, or
adiabatic) (FS) approach. They are related in  high energy regime
for special cases of the scattering amplitudes.

%-----------------------------------------------------------------
%   IMPULSE approximation
%-----------------------------------------------------------------

\subsection{The factorized impulse approximation [FIA]}

Within the impulse approximation, the interaction between the clusters
$V_{{\cal I}{\cal J}}$
is assumed to have a negligible dynamical effect on the scattering of the
projectile from the individual target subsystems
and therefore can be neglected.
The operator projectile-${\cal I}$ target
subsystem transition amplitude $\tau_{\cal I}$ is then replaced by
\be
\hat{t}_{\cal I} = v_{\cal I} + v_{\cal I} \hat{G}_0 \hat{t}_{\cal I}
\ee
where $\hat{G}_0$ contains only the kinetic energy operator $K$
\be
 \hat{G}_0 = \left( E^+ - K \right)^{-1}~~.
\ee
The transition amplitude $\hat{t}_{\cal I}$ is still a many body operator,
because the kinetic energy operator has contributions from the projectile
and all ${\cal N}$ target  subsystems.
The interaction between the target subsystems leads
to terms in 3rd order of the $\hat{t}_{\cal I}$ transition amplitude.
As we shall see, this projectile-${\cal I}$
subsystem amplitude can be reduced,  after suitable approximations,
to a free two-body amplitude evaluated at the appropriate energy.
Accepting the validity of replacing $\tau_{\cal I}$ by $\hat{t}_{\cal I}$ in
Eq.~(\ref{Tfullexp}) we obtain the multiple scattering expansion
\be
T^{\rm IA} = \sum_{\cal I} \hat{t}_{\cal I} + \sum_{\cal I}
 \hat{t}_{\cal I} \hat{G}_0 \sum_{{\cal J}
\neq {\cal I}} \hat{t}_{\cal J} + \cdots
\label{TMSexpa}
\ee
In the single scattering approximation (SA) only the first term is taken
into account.
%===========================================================================

Let us consider the scattering from subsystem ${\cal I}=2$.
Because from the dynamical point of view we want to reduce the problem to
the projectile scattering from each subsystem,
we take as relevant coordinates the relative momentum between
projectile and subsystem ${\cal I}$=2, $\vec{q}_{1,2}$,
and the relative momentum between the target subsystems, $\vec{q}_{2,3}$,
defined here as
\be
\vec{q}_{1,2} = \frac{m_2 \vec{k}_1 - m_1 \vec{k}_2}{M_{12}}
~~,~~
\vec{q}_{2,3} = \frac{m_3 \vec{k}_2 - m_2 \vec{k}_3}{M_{23}}
\label{relative1}
\ee
where  $M_{12}=m_1+m_2, M_{23}=m_2+m_3$. In the three-particle c.m. frame,
$\vec{\cal P}_t = \sum_{i=1}^3 \vec{k}_i = 0 $
($\vec{\cal P'}_t = \sum_{i=1}^3 \vec{k}\,'_i = 0 $), and
the propagator $\hat{G}_0$ is
\be
\hat{G}_0 &=& \left[E^+ - \frac{\hbar^2}{2 \mu_{12} } q^2_{12}
- \frac{\hbar^2}{2 \mu_{12,3}}  k^2_3
 \right]^{-1}
\nonumber \\
&=& \left[E^+ - \frac{\hbar^2}{2 \mu_{12} } q^2_{12}
- \frac{\hbar^2}{2 \mu_{12,3}}
\left(\vec{q}_{23} +\frac{ m_3 }{M_{23} } \vec{k}_1\right)^2
 \right]^{-1}\nonumber \\
\ee
with $\mu_{12}=m_1 m_2/M_{12}$ and $\mu_{12,3}=m_3 M_{12}/M_{123}$,
$M_{123}=m_1+m_2+m_3$.
The total energy, neglecting binding effects, is
\be
E = \frac{\hbar^2}{2 \mu_{1,23}} k_1^2 +
\frac{\hbar^2}{2 \mu_{23}}  Q_{23}^2~~,
\ee
with $\mu_{1,23}=m_1 M_{23}/M_{123}$, and $\vec{Q}_{23}$
an average relative
momentum between the target subsystems, to be specified below.
The single scattering matrix elements are
\be
& & \langle \vec{k}'_1 \phi_{23} | \hat{t}_{2} | \vec{k}_1 \phi_{23} \rangle
= \int d \vec{q}_{23}
\phi^*_{23}\left(\vec{q}_{23} - \frac{m_{3}}{ M_{{2}{3}}}
\vec{\Delta}  \right) \nonumber \\
& &  ~~~~~~~ \times
\langle  \vec{q}\,'_{1,{2}} | \hat{t}_{2} (\hat{\omega}_{1{2}})
| \vec{q}_{1,2}\rangle
  \phi_{23}\left(\vec{q}_{23} \right) ~~.
\label{fullfoldingt}
\ee
Here, $ \vec{\Delta}$ is the projectile momentum transfer
\be
\vec{\Delta} =  \vec{k}'_1 - \vec{k}_1 ~~.
\ee
The initial and final relative momenta between the projectile and the subsystem
${\cal I}=2$, $\vec{q}_{1,{2}}$ and $\vec{q}\,'_{1,{2}}$ are respectively
\be
\vec{q}_{1,{2}} &=&
\frac{\mu_{12}}{\mu_{1,23}}
\vec{k}_1 - \frac{\mu_{1{2}}}{m_{2}} \vec{q}_{{2},{3}}
= \beta_{12} \vec{k}_1 - \alpha_{12} \vec{q}_{{2},{3}} ~~,
\nonumber \\
\vec{q}\,'_{1,{2}}
&=&  \frac{\mu_{12}}{\mu_{1,23}}
\vec{k}\,'_1 - \frac{\mu_{1{2}}}{m_{2}} \vec{q}\,'_{{2},{3}}
= \beta_{12}  \vec{k}\,'_1 - \alpha_{12} \vec{q}\,'_{{2},{3}}
\nonumber \\
&=& \frac{\mu_{12}}{\mu_{1,23}}
\vec{k}\,'_1 - \frac{\mu_{1{2}}}{m_{2}} \vec{q}_{{2},{3}}
+ \frac{\mu_{1{2}}}{m_{2}} \frac{m_3}{M_{23}}   \vec{\Delta}
\nonumber \\
&=& \beta_{12}  \vec{k}\,'_1 - \alpha_{12} \vec{q}_{{2},{3}}
+ \alpha_{12} \gamma_{12} \vec{\Delta} ~~,
\label{relative3}
\ee
with  $\beta_{12}=\mu_{12}/\mu_{1,23}$,
$\alpha_{12}=\mu_{12}/m_2$, and
$\gamma_{12}= \mu_{23}/m_{2}$. We shall use these whenever a
simplified notation is required.
These new parameters satisfy $\beta_{12} + \alpha_{12} \gamma_{12} =1$,
from which it follows, together with  Eq.~(\ref{relative3}), that
the condition  $\vec{q}\,'_{1,2} - \vec{q}_{1,2} = \vec{\Delta}$
necessarily holds.
The energy parameter $\hat{\omega}_{1{2}}$ is
\be
\hat{\omega}_{1{2}} =
E -  \frac{\hbar^2} { \mu_{12,3} }
\left(\vec{q}_{23} + \frac{ m_3 }{M_{23} } \vec{k}_1\right)^2 ~~.
\label{fullfoldingE}
 \ee
The single-scattering matrix elements of Eq.~(\ref{fullfoldingt}) involve
a full folding integral of a product of a transition amplitude
and a target form factor. This integral may be quite involved.
As we shall see below, in the high energy regime, the relative momentum of
the interacting pair $\vec{q}_{23}$ can be approximated by a suitable value
$\vec{\cal Q}_{23}$. Once replaced in
Eqs.~(\ref{relative3}-\ref{fullfoldingE})
one obtains a factorized impulse approximation (FIA) expression,
\be
T^{\rm FIA}(E) =  \langle \vec{\cal Q}\,'_{12}
| \hat{t}_{2} (\omega_{12})| \vec{\cal Q}_{12}     \rangle
 \rho_{23}\left( \frac{m_3}{ M_{23} } \vec{\Delta} \right) \nonumber \\
~~~ +
 \langle \vec{\cal Q}\,'_{13}  \phi_{3}
| \hat{t}_{3} (\omega_{13})|   \phi_{3}   \vec{\cal Q}_{13}   \rangle
\rho_{23}\left( \frac{m_2}{ M_{23} } \vec{\Delta} \right) ~~,
\label{TFactorized3b}
\ee
where $\vec{\cal Q}_{1{\cal I}}$ is approximate relative momentum
projectile-subsystem ${\cal I}$, $\omega _{1{\cal I}}$
the  energy parameter and $\rho_{23}$  the target form factor
\be
\rho_{23} \left(\vec{\Delta}_1 \right)=
\int d \vec{q}_{23}
\phi^*_{23}\left(\vec{q}_{23} - \vec{\Delta}_1  \right)
 \phi_{23}\left(\vec{q}_{23} \right)
\ee
In the limit of a heavy subsystem $m_3 \gg  m_2$,
we obtain the expected limit of
 $\rho_{23}\left( \frac{m_2}{ M_{23} } \vec{\Delta} \right)=1$.
Equivalently, we can write the scattering amplitudes in terms of the
transferred momentum
$\vec{\kappa}_{12} = \vec{\cal Q}\,'_{12} - \vec{\cal Q}_{12}$
and total momentum
${\vec{\cal K}}_{12} = [\vec{\cal Q}\,'_{12} + \vec{\cal Q}_{12}]/2$:
\be
T^{\rm FIA}(E) =
\hat{t}_{2} (\omega_{12}, \vec{\kappa}_{12}, \vec{\cal K}_{12})
 \rho_{23}\left( \frac{m_3}{ M_{23} } \vec{\Delta} \right) \nonumber \\
~~~ + \hat{t}_{3} (\omega_{13},  \vec{\kappa}_{13} , \vec{\cal K}_{13}  )
\rho_{23}\left( \frac{m_2}{M_{23}} \vec{\Delta} \right) ~~.
\label{TFactorized3b-k}
\ee
The scattering amplitudes are on-shell if
$|{\vec{\cal Q}}_{1{\cal I}}|$ = $|\vec{\cal Q}\,'_{1{\cal I}}|$
= $\sqrt{2 \mu_{1,{\cal I}} \omega_{1{\cal I}}}/\hbar$.
In this case the scattering amplitudes are local.
As we shall show below, although
 $\vec{q}\,'_{1,2} - \vec{q}_{1,2} = \vec{\Delta}$,
the approximate FIA may not satisfy
$\vec{\kappa}_{12} = \vec{\Delta}$.
The scattering amplitudes are related to the transition amplitude according to
\cite{Joa87}
\be
|F(E)|^2 =\frac{(2 \pi^2)^4}{\hbar v_1} k_1^2
\frac{dk_1}{d E }  |T(E)|^2
\label{FNorelativistic-target}
\ee
for projectile target scattering, and
\be
|f_{\cal I}(\omega_{1 {\cal I} })|^2 =\frac{(2 \pi^2)^4}{\hbar v_1} k_1^2
\frac{dk_1}{d \omega_{1 {\cal I}} }
 |t_{\cal I}(\omega_{1 {\cal I} } )|^2
\label{FNorelativistic-cluster}
\ee
for projectile subsystem ${\cal I}$ scattering.
The elastic differential angular distribution
is  then evaluated  from the total scattering amplitude on-shell
\be
F^{\rm FIA}(E) = {\cal N}_{12}^{1/2}
\hat{f}_{2} (\omega_{12}, \vec{\kappa}_{12})
 \rho_{23}\left( \frac{m_3}{ M_{23} } \vec{\Delta} \right) \nonumber \\
~~~ + {\cal N}_{13}^{1/2}
 \hat{f}_{3} (\omega_{13},  \vec{\kappa}_{13}  )
\rho_{23}\left( \frac{m_2}{M_{23}} \vec{\Delta} \right) ,
\label{FFactorized3b}
\ee
where the normalization factors are
\be
{\cal N}_{1 {\cal I}} = \left[ \frac{1}{ d\omega_{1 {\cal I}}/d E }
\right]^2~~.
\ee
We now discuss several factorized impulse approximations that can be
found in the literature.
All these approximations use Eq.~(\ref{fullfoldingt}) as
starting point. The differences between the models
arise from the approximations performed to obtain a
factorized expression of the form (\ref{TFactorized3b}).

\subsubsection{The 3b on-shell FIA [Kujawski and Lambert]}

%%%%%%%%%%%%%%%%%%%%%%%%%%%%%%%%%%%%%%%%%%%%%%%%%%%%
%                 Cluster
\begin{figure}
{\par\centering \resizebox*{0.25\textwidth}{!}
{\includegraphics{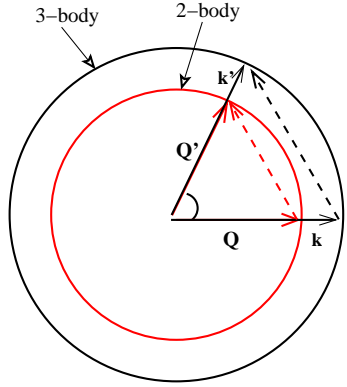}} \par}
\caption{\label{Fig:3-bodykin-OES1} (Color online)
Kinematics for the 3-body scattering in the
impulse approximation of Kujawski and Lambert (KL).
The dashed lines represent the momentum transfer for the case
of the 3-body system (dashed dark line) and
the  2-body system (dashed light line). The vectors $\vec{k} (\vec{k}\,')$
represent the initial (final) projectile momenta, and
$\vec{Q} (\vec{Q}\,')$ the initial (final) projectile-subsystem
${\cal I}$ relative momenta.
}
\end{figure}
%%%%%%%%%%%%%%%%%%%%%%%%%%%%%%%%%%%%%%%%%%%%%%%%%%%%%%%%%%%%%%%%%%

In the factorized on-shell  approximation, discussed in the work
of Kujawski and Lambert (KL) \cite{Kuj73}, the incident and outgoing
relative momenta of the subsystems $q_{23} (q'_{23})$ are
neglected by comparison with the projectile momentum.
 That is, inserting
\be
|\vec{Q}_{23}| =  0
\label{conditionOES1}
\ee
into Eq.~(\ref{relative3}-\ref{fullfoldingE}), one obtains for the relative
momenta between the projectile and subsystem ${\cal I}=2$
\be
\vec{\cal Q}_{12} =  \frac{\mu_{12}}{\mu_{1,23}}\vec{k}_1
~~, ~~ \vec{\cal Q}\,'_{12} = \frac{\mu_{12}}{\mu_{1,23}}\vec{k}\,'_1
~~,
\label{relativeOE1}
\ee
and for the energy parameter
\be
\omega_{12} &=& \frac{\hbar^2}{2 \mu_{12}} |\vec{\cal Q}_{12}|^2
 =  \frac{\mu_{1 2}}{ \mu_{1,23} } E  ~~.
\label{energyOE1-3b-12}
\ee
Similar expressions are obtained for the scattering from subsystem ${\cal I}=3$
\be
\vec{\cal Q}_{13} =  \frac{\mu_{13}}{\mu_{1,23}}\vec{k}_1
~~, ~~ \vec{\cal Q}\,'_{13} = \frac{\mu_{12}}{\mu_{1,23}}\vec{k}\,'_1
~~,
\label{relativeOE1-3}
\ee
and for the energy parameter
\be
\omega_{13} = \frac{\mu_{1 {3}}}{ \mu_{1,23} } E  ~~.
\label{energyOE1-3b-13}
\ee
The kinematics for the KL approximation is represented schematically in
Fig.\ \ref{Fig:3-bodykin-OES1}.
Clearly, in this approach the relative projectile-subsystem ${\cal I}$
momentum is a fraction of the total transfered momentum $\vec{\Delta}$
\be
\vec{\kappa}_{1{\cal I}} = \frac{\mu_{1{\cal I}}}{\mu_{1,23}}\vec{\Delta} ~~.
\ee
The single scattering terms are then
 \be
T =  \langle \frac{\mu_{12}}{\mu_{1,23}}\vec{k}\,'_1
| \hat{t}_{2} (\omega_{12})|  \frac{\mu_{12}}{\mu_{1,23}}\vec{k}_1 \rangle
 \rho_{23}\left( \frac{m_3}{ M_{23} } \vec{\Delta} \right) \nonumber \\
~~~ +
 \langle \frac{\mu_{13}}{\mu_{1,23}}\vec{k}\,'_1
| \hat{t}_{3} (\omega_{13})|  \frac{\mu_{13}}{\mu_{1,23}}\vec{k}_1 \rangle
\rho_{23}\left( \frac{m_2}{ M_{23} } \vec{\Delta} \right) ~~.
\label{TOES1-3b}
\ee
By construction from
Eqs.~(\ref{relativeOE1}-\ref{energyOE1-3b-12}),
the matrix elements of the transition amplitudes $\hat{t}_2$
and $\hat{t}_3$ are on-shell.
Equivalently one may write the transition amplitudes as a function of the
projectile momentum transfer
\be
T &=&
\hat{t}_{2} (\omega_{12}, \frac{\mu_{12}}{\mu_{1,23}}\vec{\Delta})
\rho_{23}\left( \frac{m_3}{ M_{23} } \vec{\Delta} \right) \nonumber \\
&+&
\hat{t}_{3} (\omega_{13},
\frac{\mu_{13}}{\mu_{1,23}}  \vec{\Delta})
\rho_{23}\left( \frac{m_2}{ M_{23} } \vec{\Delta} \right) ~~.
\nonumber \\
\label{TKL-II-3b}
\ee
Using Eqs.~(\ref{TOES1-3b}) and (\ref{FFactorized3b}),
the total scattering amplitude $F$
can be written in terms of the scattering amplitudes for each subsystem
$f_{\cal I}$, with normalization coefficients
\be
{\cal N}_{1{\cal I}} =
\left( \frac{\mu_{1,23}}{\mu_{1{\cal I}}}    \right)^2   ~~~,~~~
{\cal I} = 1,2 ~~.
\label{coeff1}
\ee

\smallskip

\subsubsection{The 3b on-shell FIA [Rihan]}

The optimal factorized approximation discussed in the work
of Rihan \cite{Rih77,Rih96} was formulated for a three body problem
in the context of the multiple scattering expansion of the optical potential.
It was to our knowledge never applied to a specific scattering problem.
In this approximation, the relative momentum between
the two subsystems $\vec{q}_{23}$ is taken as the mid-point
where the product of the wave functions peaks in Eq. (\ref{fullfoldingt}),
that is,
\be
\vec{Q}_{23} =  \frac{1}{2} \frac{m_3}{M_{23}} \vec{\Delta} ~~.
\ee
Substituting this into Eq.~(\ref{relative3}), the relative
momenta are
\be
\vec{\cal Q}_{12} &=&  \beta_{12}  \vec{k}_1
-   \frac{\alpha_{12} \gamma_{12}}{2}   \vec{\Delta}
\nonumber \\
\vec{\cal Q}\,'_{12} &=& \beta_{12} \vec{k}\,'_1
+ \frac{\alpha_{12} \gamma_{12}}{2} \vec{\Delta} ~~,
\label{relative_k_OFAC3b}
\ee
and the energy parameter is
\be
\omega_{12} = \frac{\hbar^2}{2 \mu_{12}} |\vec{\cal Q}_{12}|^2
=    \widehat{\cal C}_{12} \frac{\mu_{1,2}}{\mu_{1,23}}   E
\label{EnergyOFAC3b}
\ee
with
\be
\widehat{\cal C}_{12} = \left[ \frac{\mu_{1,23}}{\mu_{12}}\right]^2
[1 + {\cal C}_{12} + {\cal C}_{12}\, \cos\theta ] ~~.
\ee
Here, $\theta$ is the projectile-target scattering angle and
\be
{\cal C}_{12} = - \alpha_{12}  \gamma_{12}
 + \frac{1}{2}\alpha_{12}^2 \gamma_{12}^2 ~~.
\ee
For small scattering angles $\widehat{\cal C}_{12} \sim 1$. Similar expressions
can be obtained for the scattering from subsystem ${\cal I}=3$.

\begin{figure}
{\par\centering \resizebox*{0.25\textwidth}{!}
{\includegraphics{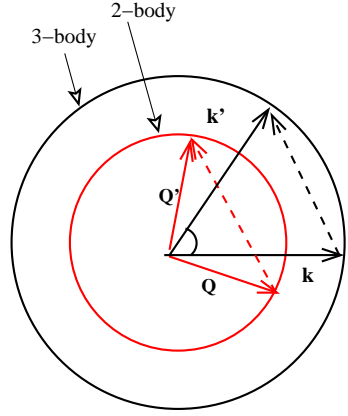}} \par}
\caption{\label{Fig:rihan}(Color online)
 Kinematics for the 3-body scattering in
optimal factorization impulse approximation of Rihan.
The lines and vectors have the same meaning as
in Fig.~\ref{Fig:3-bodykin-OES1}.
}
\end{figure}

The kinematics for the Rihan approximation is represented schematically
in Fig.~\ref{Fig:rihan}.
Within this approximation, using $\beta_{12} + \alpha_{12} \gamma_{12}=1$,
it follows that
\be
\vec{\kappa}_{1{\cal I}} = \vec{\Delta}
\ee
The single-scattering matrix elements are
\begin{widetext}
\be
T &=&
\langle \beta_{12} \vec{k}\,'_1
+ \frac{\alpha_{12} \gamma_{12}}{2} \vec{\Delta}
| \hat{t}_{2} ({\omega}_{12})|
 \beta_{12}  \vec{k}_1
-   \frac{\alpha_{12} \gamma_{12}}{2}   \vec{\Delta}    \rangle
%\times
\rho_{23}\left( \frac{m_3}{ M_{23} } \vec{\Delta} \right)
\nonumber \\
&+&
\langle \beta_{13} \vec{k}\,'_1
+ \frac{\alpha_{13} \gamma_{13}}{2} \vec{\Delta}
| \hat{t}_{3} ({\omega}_{13})|
 \beta_{13}  \vec{k}_1
-   \frac{\alpha_{13} \gamma_{13}}{2}   \vec{\Delta} \rangle
%\times
\rho_{23}\left( \frac{m_2}{ M_{23} } \vec{\Delta} \right)
~~.
\label{TOFAC3b}
\ee
\end{widetext}
Equivalently one may write the transition amplitude in terms of the
projectile momentum transfer
\begin{widetext}
\be
T &=& \hat{t}_{2} \left ({\omega}_{12},  \vec{\Delta} \right)
\rho_{23}\left( \frac{m_3}{ M_{23} } \vec{\Delta} \right) +
\hat{t}_{3} \left({\omega}_{13},  \vec{\Delta} \right)
\rho_{23}\left( \frac{m_2}{ M_{23} } \vec{\Delta} \right)
~~.
\ee
\end{widetext}
By construction from Eq.~(\ref{relative_k_OFAC3b})
and Eq.~(\ref{EnergyOFAC3b}),
the matrix elements are on-shell.

%%%%%%%%%%%%%%%%%%%%%%%%%%%%%%%%%%%%%%%%%%%%%%%%%%%%%%%%%%%%%%%%%%%%%%%%%%
%
%%%%%%%%%%%%%%%%%%%%%%%%%%%%%%%%%%%%%%%%%%%%%%%%%%%%%%%%%%%%%%%%%%%%%%%%%%

Making use of Eqs.~(\ref{TOES1-3b}) and (\ref{FFactorized3b}),
% -\ref{normscat})
the total scattering amplitude $F$
can be written in terms of the scattering amplitudes for each subsystem
$f_{\cal I}$, where the normalization coefficients are
\be
{\cal N}_{1{\cal I}}= \left( \frac{\mu_{1,23}}{\mu_{1{\cal I}}} \right)^2
\left( \frac{1}{ \widehat{\cal C}_{1{\cal I}}}\right)^2    ~~~,~~~
{\cal I} = 1,2 ~~.
\label{coeffOF}
\ee
We note that for small scattering angles the energy parameters
reduce to Eqs.~(\ref{energyOE1-3b-13}) and (\ref{energyOE1-3b-12})
and the normalization coefficients to Eq.~(\ref{coeff1}).

%-----------------------------------------------------------------------------
% CHEW
%-----------------------------------------------------------------------------

\subsubsection{The 3b on-shell FIA [Chew]}

\begin{figure}
{\par\centering \resizebox*{0.25\textwidth}{!}
{\includegraphics{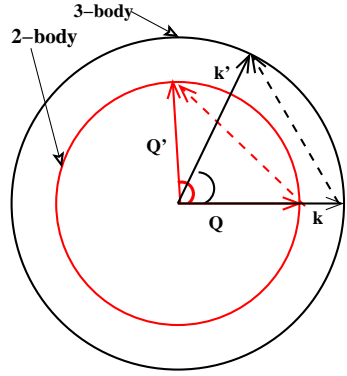}} \par}
\caption{\label{Fig:3-bodykin-chew}(Color online)
Kinematics for the 3-body scattering in the
on-shell impulse approximation of Chew.
The lines and vectors have the same meaning
as in Fig.~\ref{Fig:3-bodykin-OES1}.
}
\end{figure}

For completeness we discuss now the approximation discussed in
\cite{Chew51,Chew50}, although no practical application will be
made here.  Within this approximation, the relative incident
momenta between the subsystems $q_{23}$ is again neglected when
compared with the incident projectile momentum. That is, setting
$Q_{23} = 0 $ one gets
\be
\vec{\cal Q}_{12} =  \frac{\mu_{12}}{\mu_{1,23}}\vec{k}_1~~.
\label{Chewqq}
\ee
In addition, it is also assumed that
\be
 |\vec{\cal Q}\,'_{12}| =  |\vec{\cal Q}_{12}|
~~, ~~ |\vec{\kappa}| = |\vec{\Delta}| ~ .
\label{relativeOE2}
\ee
The kinematics in the Chew approximation is represented schematically in
Fig.~\ref{Fig:3-bodykin-chew}.

The energy parameters, $\omega_{12}$ and  $\omega_{13}$
are given by Eq.~(\ref{energyOE1-3b-12})
and Eq.~(\ref{energyOE1-3b-13}) respectively.
By construction from these equations the matrix elements are on the
energy shell.
The normalization coefficients for the angular scattering amplitudes
are  identical to those derived in the KL approximation Eq. (\ref{coeff1}).

%%%%%%%%%%%%%%%%%%%%%%%%%%%%%%%%%%%%%%%%%%%%%%%%%%%%
%
%%%%%%%%%%%%%%%%%%%%%%%%%%%%%%%%%%%%%%%%%%%%%%%%%%%%%%%%%%%%%%%%%%%%%

\subsubsection{The 3b on-shell FIA [Crespo and Johnson]}

In this approach, followed in the most current applications of
MST \cite{Cres99,Cres02,CresRNB},
the initial internal relative momentum $q_{23}$  is neglected in
the transition matrix elements,
and thus Eq. ({\ref{Chewqq}) is satisfied. In addition,
\be
 \vec{\cal Q}\,'_{12} = \vec{\cal Q}_{12} + \vec{\Delta}
\ee
The energy parameters, $\omega_{12}$ and  $\omega_{13}$
are given by Eq.~(\ref{energyOE1-3b-12})
and Eq.~(\ref{energyOE1-3b-13}) respectively.

The single-scattering terms can then be written as a
function of the transition amplitude for proton-subsystem scattering
and the density for the subsystem $\rho_{23}$,
\be
T =  \langle \frac{\mu_{12}}{\mu_{1,23}}\vec{k}\,_1 + \vec{\Delta}
| \hat{t}_{2} (\omega_{12})|  \frac{\mu_{12}}{\mu_{1,23}}\vec{k}_1 \rangle
 \rho_{23}\left( \frac{m_3}{ M_{23} } \vec{\Delta} \right) \nonumber \\
~~~ +
 \langle \frac{\mu_{13}}{\mu_{1,23}}\vec{k}_1 + \vec{\Delta}
| \hat{t}_{3} (\omega_{13})|  \frac{\mu_{13}}{\mu_{1,23}}\vec{k}_1 \rangle
\rho_{23}\left( \frac{m_2}{ M_{23} } \vec{\Delta} \right)
\nonumber \\
\label{TOFF-I-3b}
\ee
or equivalently
\be
T =  \hat{t}_{2} (\omega_{12},\Delta,{\cal K}_{12},\phi)
 \rho_{23}\left( \frac{m_3}{ M_{23} } \vec{\Delta} \right) \nonumber \\
~~~ +
\hat{t}_{3} (\omega_{13},\Delta,{\cal K}_{13},\phi)
\rho_{23}\left( \frac{m_2}{ M_{23} } \vec{\Delta} \right) ~~,
\nonumber \\
\label{TOFF-II-3b}
\ee
with $\phi$ the angle between $\Delta$ and ${\cal K}$. On the energy shell,
$\phi=\pi/2$.
The matrix elements of the transition amplitude are half on the energy shell
since ${\cal Q}_{12} = \sqrt{2 \mu_{12} \omega_{12}}/\hbar$,
but ${\cal Q}\,'_{12} \neq {\cal Q}_{12}$,
as represented schematicaly in Fig.~\ref{Fig:3-bodykin-OFFES}.

%%%%%%%%%%%%%%%%%%%%%%%%%%%%%%%%%%%%%%%%%%%%%%%%%%%%
%
\begin{figure}
{\par\centering \resizebox*{0.25\textwidth}{!}
{\includegraphics{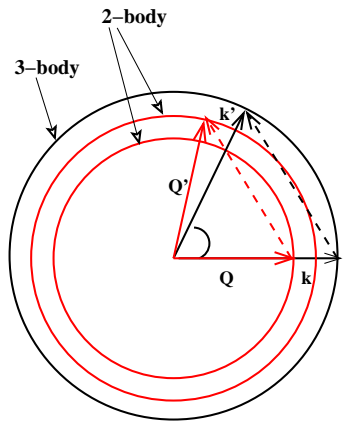}} \par}
\caption{\label{Fig:3-bodykin-OFFES}(Color online)
Kinematics for the 3-body scattering in the
off-energy shell, OFFES, impulse approximation.
The lines and vectors have the same meaning as in
Fig.~\ref{Fig:3-bodykin-OES1}.
}
\end{figure}
%%%%%%%%%%%%%%%%%%%%%%%%%%%%%%%%%%%%%%%%%%%%%%%%%%%%%%%%%%%%%%%%%%%%

This approach is however impractical if one has only on-shell
scattering amplitudes for the scattering from the
subsystems. This is the case, for example, when these amplitudes are obtained
by fitting scattering data.
We note that for small scattering angles, the scattering angle for
projectile subsystem ${\cal I}=2$ scattering, $\theta_{12}$, satisfies
\be
\theta_{1,2}
\sim  \frac{ \mu_{1(23)} }{\mu_{1 {2}}} \theta  ~~.
\ee
This means  that the range of physical momentum
transfers over which the on-shell $t_{12}$ transition amplitude
is defined is smaller than the range of momentum transfers accessible
in for the scattering from the 3-body system.

If the dependence of the transition amplitude
$\hat{t}_{\cal I} (\omega_{1{\cal I}},\Delta,{\cal K}_{1{\cal I}},\phi)$
in Eq.  (\ref{TOFF-II-3b})
on the total momentum is assumed to be small
then we can replace the matrix elements by their on-shell values
at the appropriate energy. On the energy shell, the total momentum ${\cal K}$
is related to the momentum transfer $\Delta$,
and $\phi=\pi/2$.
In most current multiple scattering applications
\cite{Cres99,Cres02,CresRNB} the on-shell approximation is made
within a partial wave expansion of the transition amplitude, but
the angular direction is kept constant.
We refer to this factorized impulse approximation approach as
projected on-energy shell [POES].

Making use of Eq.~(\ref{TOFF-II-3b}) and Eq.~(\ref{FFactorized3b})
the total scattering amplitude $F$
can be written in terms of the scattering amplitudes for each subsystem
$f_{\cal I}$, where the normalization coefficients are given as
in Eq.~(\ref{coeff1}).
%-----------------------------------------------------------------
%   ADIABATIC or FSA approximation
%-----------------------------------------------------------------

\subsection{The Fixed scatterer or adiabatic approximation [FSA]}

Within the fixed scatterer or adiabatic approximation,
the internal Hamiltonian between the clusters
is taken to a constant $\bar{H}$, that is, the operator projectile-${\cal I}$
target subsystem transition amplitude $\tau_{\cal I}$ is replaced by
\be
\tilde{t}_{\cal I} = v_{\cal I} + v_{\cal I} \tilde{G}_0 \tilde{t}_{\cal I}
\ee
where $\tilde{G}_0$
\be
 \tilde{G}_0 = \left( E^+ - K_1 - \bar{H}  \right)^{-1}~~.
\ee
Even after this simplification, the solution of this problem involves
the solution of a system of coupled equations, which
in practice requires truncation of the angular momenta in order to make
the problem solvable \cite{Chr97}.
A notable simplification of the problem is achieved in general when one
neglects the interaction of the projectile with all the fragments
except the core \cite{Ron97b, Cres99}, the
so called core-recoil model. In this
particular case, the few-body problem can be exactly solved, leading
to a simple factorized expression,  the product of the transition amplitude
for the scattering from the core evaluated at the appropriate energy
and a structure form factor.

Within the FSA framework for the case where the interaction between the
projectile and fragment 3 is neglected
the total transition amplitude takes the form,
\be
T^{\rm FSA} =  \langle \vec{k}\,'_{1}
| \tilde{t}_{2} (E)| \vec{k}_{1}     \rangle
 \rho_{23}\left( \frac{m_3}{ M_{23} } \vec{\Delta} \right)  ~~.
\label{FSA-3b-ron}
\ee
 We note that in the FSA approach, the energy parameter is
$E=\hbar k_1^2/2 \mu_{1,23}$ and thus,
a distinctive feature of the FSA is that the two-body amplitude
is calculated with the reduced mass $\mu_{1,23}$, instead of the
reduced mass appropriate for the two-body scattering
which appears in the FIA scattering amplitude.
In the general case,
i.e., when the interaction with the two target subsystems is considered,
one can perform a multiple scattering expansion of the few-body
amplitude in terms of the individual T-matrices
$\tilde{t}_{\cal I}$ for the fragments which, in leading order, yields
the factorized expression,
\be
T^{\rm FSA} =  \langle \vec{k}\,'_{1}
| \tilde{t}_{2} (E)| \vec{k}_{1}     \rangle
 \rho_{23}\left( \frac{m_3}{ M_{23} } \vec{\Delta} \right) \nonumber \\
~~~ +
 \langle \vec{k}\,'_{1}
| \tilde{t}_{3} (E)| \vec{k}_{1}   \rangle
\rho_{23}\left( \frac{m_2}{ M_{23} } \vec{\Delta} \right) ~~,
\label{FSA-3b}
\ee
In principle, the fixed scatterer approximation is conceptually
different from the factorized impulse approximation, in any of the versions
discussed above. However,we shall show further down that in the
high energy limit, the calculated elastic scattering observables
using the FSA and the FIA (POES or Rihan) are essentially identical.
The formal relation between the FSA and FIA and their relation with
the Glauber theory will be explored elsewhere.

\section{The single scattering 4-body problem}

We consider now the case of proton scattering from a nucleus assumed
to be well described by a three body model (for example $^{11}$Li),
as a core (here labeled as subsystem 4) and two
valence weakly bound  systems (subsystems 2 and 3).
Neglecting core excitation, the  target wave function  is
\be
\Phi(\vec{r},\vec{R},{\xi}_4)
= \left[ \phi_{23,4}(\vec{r},\vec{R}) \otimes
 \varphi_{4}(\xi_4) \right] ~~, \label{FBwf3b}
\ee
where $\varphi_{4 }(\xi_4)$ is the core internal wave function
and $\phi_{23,4}(\vec{r},\vec{R})$
the  three body valence wave function  relative to the core. The
internal degrees of freedom of the core are denoted globally
as $\xi_4$.
The total transition amplitude is then
\be
\label{eq:ssa-4b}
T^{IA} &=&  \langle \vec{k}\,'_1 \Phi  | \hat{t}_{12} | \vec{k}_1 \Phi \rangle
+ \langle \vec{k}\,'_1 \Phi |  \hat{t}_{13} | \vec{k}_1 \Phi \rangle
+  \langle \vec{k}\,_1 \Phi | \hat{t}_{14}  | \vec{k}_1 \Phi \rangle
\nonumber \\
&+& \cdots
\ee
where within the single scattering approximation only the first three terms
are taken into account.
%%%%%%%%%%%%%%%%%%%%%%%%%%%%%%%%%%%%%%%%%%%%%%%%%%%%%%%%%%%%%%%%%%%%%%%%%%%
We first consider the situation where particle 1 scatters from
one of the valence systems, named  subsystem 2.
The relevant Jacobian coordinates are
\be
\vec{q}_{1,2}   &=& \frac{m_2 \vec{k}_1 - m_1 \vec{k}_2}{M_{12}} \\
\vec{q}_{2,3}   &=& \frac{m_3 \vec{k}_2 - m_2 \vec{k}_3}{M_{23}}  \\
\vec{q}_{23,4}  &=& \frac{m_4 \vec{P}_{23} - M_{23} \vec{k}_4}{M_{234}}
~~,
\ee
where  $ \vec{P}_{12} = \vec{k}_1 + \vec{k}_2 $,
$ \vec{P}_{34} = \vec{k}_3 + \vec{k}_4 $, etc.
In the center of mass of the total four-body system,
 $\vec{\cal P}_t = \sum_{i=1}^4 \vec{k}_i = 0$.
The intermediate states propagator is given by
\be
\hat{G}_0 = \left(E^+ - \frac{\hbar^2}{2 \mu_{12}} \vec{q}^{\,2}_{\,12}
- \frac{\hbar^2}{2 \mu_{34}} \vec{q}^{\,2}_{\,34}
- \frac{\hbar^2}{2 \mu_{12,34}} \vec{q}^{\,2}_{\,12,34}\right)^{-1}
\ee
with
\be
\vec{q}_{12,34} &=&
\frac{ M_{34} \vec{P}_{12} - M_{12}  \vec{P}_{34}}{ M_{1234} } \\
\vec{q}_{3,4} &=& \frac{m_4 \vec{k}_3 - m_3 \vec{k}_4}{M_{34}}
\ee
and $\mu_{12}$, $\mu_{34}$, $\mu_{12,34} = M_{12} M_{34} / M_{1234}$
the appropriate reduced masses.
Using the fact that the operator $\hat{t}_{12}$ is independent of the spatial
variables of the core, the single scattering matrix elements are given by
\begin{widetext}
\be
%& &
 \langle \vec{k}'_1 \phi_{23,4} | \hat{t}_{2} | \vec{k}_1 \phi_{23,4} \rangle =
\int \int  d \vec{q}_{23} \, d \vec{q}_{23,4}
\phi^*_{23,4}\left(
\vec{q}_{23} - \frac{m_3}{M_{23}}\vec{\Delta},
\vec{q}_{23,4} - \frac{m_4}{M_{234}}\vec{\Delta}\right)
\langle  \vec{q}\,'_{1,{2}} | \hat{t}_{2} (\hat{\omega}_{1{2}})
| \vec{q}_{1,2}\rangle
  \phi_{23,4}\left(  \vec{q}_{23}, \vec{q}_{23,4}    \right)
\label{fullfoldingt4b}
\ee
\end{widetext}
The initial (and final) relative momentum between the projectile
and the subsystem
${\cal I}=2$, $\vec{q}_{1,{2}}$ ($\vec{q}\,'_{1,{2}}$) are respectively
\be
\vec{q}_{1,{2}} &=&
  \frac{\mu_{12}}{\mu_{1,234}} \vec{k}_1
- \frac{\mu_{12}}{m_2}
\left [ \vec{q}_{23} + \frac{m_2}{M_{23}} \vec{q}_{23,4} \right]
\nonumber \\
%%%%%%%%%%%%%%%%%%%%%%%%%%%%%%%%%%%%%%%%%%%%%%%%%%%%%%
\vec{q}\,'_{1,{2}}  &=&
  \frac{\mu_{12}}{\mu_{1,234}} \vec{k}\,'_1
 -   \frac{\mu_{12}}{m_2}
\left [ \vec{q}\,'_{23} +  \frac{m_2}{M_{23}} \vec{q}\,'_{23,4} \right]
\nonumber \\
%%%%%%%%%%%%%%%%%%%%%%%%%%%%%%
 &=&
  \frac{\mu_{12}}{\mu_{1,234}} \vec{k}\,'_1
 -   \frac{\mu_{12}}{m_2}
\left [ \vec{q}_{23} +  \frac{m_2}{M_{23}} \vec{q}_{23,4} \right]
\nonumber \\
& -&   \frac{\mu_{12}}{m_2}
\left [ \frac{m_3}{M_{23}} +  \frac{m_2}{M_{23}} \frac{m_4}{M_{234}} \right]
\vec{\Delta} ~~,
\label{relative4}
\ee
with $\alpha_{12}=\mu_{12}/m_2$  defined as in the 3-body case.
The energy parameter $\hat{\omega}_{1{2}}$ is
\be
\hat{\omega}_{12} &=& E - \frac{\hbar^2}{2 \mu_{34}}
\left[  \frac{m_3 M_{234}}{M_{23}M_{34}}  \vec{q}_{23,4}
-   \frac{m_4}{M_{34}} \vec{q}_{23} \right]^{\,2}
\nonumber \\
&-& \frac{\hbar^2}{2 \mu_{12,34}}
\left[  \frac{M_{34}}{M_{234}} \vec{k}_1
+  \frac{m_2}{M_{23}} \vec{q}_{23,4}
+  \vec{q}_{23}\right]^{\,2} \nonumber \\
\label{fullfoldingE4b}
 \ee
We note that expressions Eq.~(\ref{relative4}) and Eq.~(\ref{fullfoldingE4b})
reduce to the 3-body problem in the limit where we take
$m_4=0$.
%%%%%%%%%%%%%%%%%%%%%%%%%%%%%%%%%%%%%%%%%%%%%%%%%%%%%%%%%%%%%%%%%%%%%%%%%%%%%%

We now consider the situation where particle 1 scatters from
the core named  subsystem 4.
The relevant Jacobian coordinates are $\vec{q}_{2,3},\vec{q}_{23,4}$ and
\be
\vec{q}_{1,4} &=& \frac{m_4 \vec{k}_1 - m_1 \vec{k}_4}{M_{14}} ~~.
\ee
The intermediate states propagator  is given by
\be
\hat{G}_0 = \left(E^+ - \frac{\hbar^2}{2 \mu_{14}} \vec{q}^{\,2}_{\,14}
- \frac{\hbar^2}{2 \mu_{23}} \vec{q}^{\,2}_{\,23}
- \frac{\hbar^2}{2 \mu_{14,23}} \vec{q}^{\,2}_{\,14,23}\right)^{-1}
~~,
\ee
with
\be
\vec{q}_{14,23} =
\frac{ M_{23} \vec{P}_{14} - M_{14}  \vec{P}_{23}}{ M_{1234} }
\ee
and
$\mu_{14}$,  $\mu_{23}$, $\mu_{14,23} = M_{14} M_{23} / M_{1234}$
the appropriate reduced masses.
The single scattering matrix elements for the scattering from the core
\begin{widetext}
\be
& & \langle \vec{k}'_1 \phi_{23} \phi_4| \hat{t}_{4}
 |\varphi_4  \phi_{23,4}  \vec{k}_1  \rangle =
\int \int  d \vec{q}_{23} \, d \vec{q}_{23,4}
\phi^*_{23,4}\left( \vec{q}_{23},
\vec{q}_{23,4} + \frac{M_{23}}{M_{234}}\vec{\Delta}\right)
\langle  \vec{q}\,'_{1,{4}} \phi_4   | \hat{t}_{4} (\hat{\omega}_{1{4}})
| \varphi_4  \vec{q}_{1,4}\rangle
  \phi_{23,4}\left(  \vec{q}_{23}, \vec{q}_{23,4}    \right)
\label{fullfoldingt-core4b}
\ee
\end{widetext}
The initial (and final) relative momentum between the projectile
and the subsystem ${\cal I}=4$, $\vec{q}_{1,4}$ ($\vec{q}\,'_{1,4}$)
are respectively
\be
\vec{q}_{1,4} &=&
\frac{\mu_{14}}{\mu_{1,234}} \vec{k}_1
+  \frac{\mu_{14}}{m_4}  \vec{q}_{23,4} ~~,
\nonumber \\
\vec{q}\,'_{1,{4}}  &=& \frac{\mu_{14}}{\mu_{1,234}} \vec{k}\,'_1
+    \frac{\mu_{14}}{m_4}  \vec{q}\,'_{23,4} ~~.
\label{relative-core4b}
\ee
The energy parameter $\hat{\omega}_{14}$ is
\be
\omega_{14} = E - \frac{\hbar^2}{2 \mu_{23}}q_{23}^{\,2}
- \frac{\hbar^2}{2 \mu_{14,23}}
\left[ \frac{M_{23}}{M_{234}}  \vec{k}_1 - \vec{q}_{23,4}\right]^{\,2}
~~.
\label{fullfoldingE-core4b}
 \ee
In the limit where we take $m_3=0$, these equations reduce to the 3-body case.
%%%%%%%%%%%%%%%%%%%%%%%%%%%%%%%%%%%%%%%%%%%%%%%%%%%%%%%%%%%%%%%%%%%
As in the 3-body case, it is desirable to obtain after suitable approximations
a factorized expression of the transition amplitude matrix elements and the
target form factor
\be
\rho_{23,4}(\vec{\Delta}_{1},\vec{\Delta}_{2}) &=&
\int d\vec{Q}_1 d\vec{Q}_2 \,  \phi^*_{23,4}
(\vec{Q}_1 + \vec{\Delta}_1, \vec{Q}_2  + \vec{\Delta}_2) \nonumber \\
& \times &  \phi_{23,4}(\vec{Q}_1,\vec{Q}_2 )~~ \label{rho2} ~~.
\ee
In here $\phi_{23,4}(\vec{Q}_1,\vec{Q}_2)$ is the
Fourier transform of the wave function of the two body valence system relative
to the core $\phi_{_{23,4}}(\vec{r},\vec{R})$.
The factorized impulse approximation expression for the 4-body case is
\be
T^{\rm FIA} &=&  \langle \vec{\cal Q}\,'_{12}
| \hat{t}_{2} (\omega_{12})| \vec{\cal Q}_{12}     \rangle
 \rho_{23,4}\left( \frac{m_3}{ M_{23} } \vec{\Delta},
\frac{m_4}{ M_{234} } \vec{\Delta} \right) \nonumber \\
 &+& \langle \vec{\cal Q}\,'_{13}
| \hat{t}_{3} (\omega_{13})| \vec{\cal Q}_{13}   \rangle
\rho_{23,4}\left( \frac{m_2}{ M_{23} } \vec{\Delta},
\frac{m_4}{ M_{234} } \vec{\Delta} \right)
\nonumber \\
 &+& \langle   \vec{\cal Q}\,'_{14}   \phi_4
| \hat{t}_{4} (\omega_{14})| \phi_4   \vec{\cal Q}_{14}   \rangle
\rho_{23,4}\left( 0,
\frac{M_{23}}{ M_{234} } \vec{\Delta} \right)
\nonumber \\
\label{TFactorized4b}
\ee
where $\vec{\cal Q}$ and $\omega$ are the appropriate relative momenta
and energy parameter respectively to be discussed bellow,
which are a generalization of the 3 body case.

\subsubsection{The 4b on-shell FIA [Kujawski and Lambert]}

Within this approximation, the relative momenta between the
subsystems $q_{23}$ and $q_{23,4}$
are neglected when compared with the incident projectile momenta in the
matrix elements of the projectile subsystem transition matrix elements
whenever they appear, that is,
\be
|\vec{Q}_{23}| = |\vec{Q}\,'_{23}| =  0 \nonumber \\
|\vec{Q}_{23,4}| = |\vec{Q}\,'_{23,4}| =  0 ~~.
\label{conditionOES1-4b}
\ee
From Eq.~(\ref{relative4}) one obtains for the relative
momenta between the projectile and subsystem ${\cal I}=2$
\be
\vec{\cal Q}_{12} =  \frac{\mu_{12}}{\mu_{1,234}}\vec{k}_1
~~, ~~ \vec{\cal Q}\,'_{12} = \frac{\mu_{12}}{\mu_{1,234}}\vec{k}\,'_1
\label{relativeOE1-4b}
\ee
and for the energy parameter
\be
\omega_{12} = \frac{\mu_{1 {2}}}{ \mu_{1,234} } E
\label{energyOE1-4b}
\ee
and similarly for the scattering from subsystems ${\cal I}=3, 4$.
The single scattering terms can then be written as
\be
& & T =    \langle  \frac{\mu_{14}}{\mu_{1,234}} \vec{k}\,'_1
| \hat{t}_{4} (\omega_{14})|   \frac{\mu_{14}}{\mu_{1,234}}  \vec{k}_1 \rangle
\rho_{23,4}\left( 0, \frac{m_{23}}{ M_{234} } \vec{\Delta} \right) \nonumber \\
 & + & \langle \frac{\mu_{12}}{\mu_{1,234}}\vec{k}\,'_1
| \hat{t}_{2} (\omega_{12})|  \frac{\mu_{12}}{\mu_{1,234}}\vec{k}_1 \rangle
 \rho_{23,4}\left( \frac{m_3}{ M_{23} } \vec{\Delta},
\frac{m_4}{ M_{234} } \vec{\Delta} \right) \nonumber \\
 & + &
 \langle \frac{\mu_{13}}{\mu_{1,234}}\vec{k}\,'_1
| \hat{t}_{3} (\omega_{13})|  \frac{\mu_{13}}{\mu_{1,234}}\vec{k}_1 \rangle
\rho_{23,4}\left( \frac{m_2}{ M_{23} } \vec{\Delta},
\frac{m_4}{ M_{234} } \vec{\Delta} \right) ~~.\nonumber \\
\label{TOES1-4b}
\ee
By construction, from Eqs.~(\ref{relativeOE1-4b}-\ref{energyOE1-4b})
the matrix elements of the transition amplitudes are on-shell.

As in the three-body case the total scattering amplitude, $F$,
can be written in terms of the scattering amplitudes for each subsystem
$f_{\cal I}$, where the normalization coefficients are
\be
{\cal N}_{1{\cal I}} = \left( \frac{\mu_{1,234}}{\mu_{1{\cal I}}} \right)^2
 ~,~ {\cal I} = 2, 3, 4
\ee

\subsubsection{The 4b on-shell FIA [Rihan]}

The extension of the optimal factorized approximation discussed in the work
of Rihan \cite{Rih77,Rih96}  to the four-body problem is straightforward.
The relative momentum between
the  subsystems is taken to the mid-point value where the product of
the wave function peaks. For the scattering from subsystem
${\cal I}=2$  this yields
\be
\vec{Q}_{23} &=&  \frac{1}{2} \frac{m_3}{M_{23}} \vec{\Delta}
\nonumber \\
\vec{Q}_{23,4} &=&  \frac{1}{2} \frac{m_4}{M_{234}} \vec{\Delta}~~,
\ee
which leads to
\be
\vec{\cal Q}_{1,{2}} &=&
   \hat{\beta}_{12} \vec{k}_1
-   \frac{\alpha_{12} \hat{\gamma}_{12} }{2}  \vec{\Delta}~~,
\ee
with $\alpha_{12}$ defined as in the three-body case and
$\hat{\beta}_{12} = \mu_{12}/\mu_{1,234} $,
$\hat{\gamma}_{12}=M_{34}/M_{234}$.
Note that these expressions reduce to the 3-body case in the limit
where we take $m_4=0$.
Since this approximation involves on-shell matrix elements
the energy parameter can be evaluated as
\be
\omega_{12} =   \frac{\hbar^2}{2 \mu_{12}} |\vec{{\cal Q}}_{12}|^2
 =  E \frac{\mu_{12}}{\mu_{1,234}} \widehat{\cal C}_{12}
\ee
with
\be
\widehat{\cal C}_{12} =
\left[\frac{\mu_{1,234}}{\mu_{12}} \right]^2
[1 + {\cal C}_{12} + {\cal C}_{12}\, \cos\theta] ~~,
\label{eq:wC12_4b}
\ee
where $\theta$ is the scattering angle and
\be
{\cal C}_{12} &=& - \alpha_{12}  \hat\gamma_{12}
 + \frac{1}{2}\alpha_{12}^2 \hat{\gamma}_{12}^2 ~~.
\label{eq:C12_4b}
\ee
The normalization factor is formally identical to
the three-body case, Eq.~(\ref{coeffOF}), with
$\widehat{\cal C}_{12}$ replaced by Eq.~(\ref{eq:wC12_4b}).
Similar expressions can be derived for the case of the scattering from
subsystem ${\cal I}=3$. In the case of the scattering from
subsystem ${\cal I}=4$, the optimal approximation prescribes
\be
\vec{Q}_{23} &=& 0
\nonumber \\
\vec{Q}_{23,4} &=&  - \frac{1}{2} \frac{M_{23}}{M_{234}} \vec{\Delta}
\ee
which leads to
\be
\vec{\cal Q}_{1,4} &=&
\frac{\mu_{14}}{\mu_{1,234}} \vec{k}_1
-  \frac{\mu_{14}}{m_4} \frac{1}{2} \frac{M_{23}}{M_{234}} \vec{\Delta}
\nonumber \\
 &=&  \hat{\beta}_{14} \vec{k}_1
- \frac{\alpha_{14} \hat{\gamma}_{14}}{2}{\vec{\Delta}}
\label{relative-core4bopt}
\ee
with $\hat{\beta}_{14} = \frac{\mu_{14}}{\mu_{1,234}}$,
$\hat{\gamma}_{14} = M_{23}/M_{234}$
and for the energy parameter,
\be
\omega_{14} =  \frac{\hbar^2}{2 \mu_{14}}{\cal Q}_{14}^2
= E \frac{\mu_{14}}{\mu_{1,234}} \widehat{\cal C}_{14}
\ee
with
\be
\widehat{\cal C}_{14} =
\left[ \frac{\mu_{1,234}}{\mu_{14}} \right]^2
\left[
1 + {\cal C}_{14} + {\cal C}_{14}\, \cos\theta \right]
\ee
where
\be
{\cal C}_{14} = - \alpha_{14}  \hat{\gamma}_{14}
 + \frac{1}{2}\alpha_{14}^2 \hat{\gamma}_{14}^2 ~~.
\ee

\subsubsection{The 4b on-shell FIA [Crespo and Johnson]}

In this approach the initial relative momenta
$q_{23}, q_{23,4}$  are neglected in
the transition matrix elements momentum transfer.
For the scattering of subsystem ${\cal I}=2$
\be
\vec{\cal Q}_{12} =  \frac{\mu_{12}}{\mu_{1,234}}\vec{k}_1
~~, ~~ \vec{\cal Q}\,'_{12} = \vec{\cal Q}_{12} + \vec{\Delta}
\ee
and similarly for the scattering of the other subsystems.

\section{The $^6$H\protect\lowercase{e} structure model} \label{sec6he}

The  $^6$He is here described as a
three-body system $n$+$n$+$^4$He.
The bound  wave functions are obtained by solving the Schr\"odinger equation
in hyperspherical coordinates. We consider  two structure models
defined in terms of different effective 3-body (3B) potentials,
which are introduced to overcome the underbinding caused by the other closed
channels, most important of which the $t$+$t$ breakup.
In both models the $n$-$^4$He potential is taken
from Ref.~\cite{Bang79,Tho00}, and use
the GPT $NN$ potential \cite{gpt} with spin-orbit
and tensor components.
In the first model (R5) of \cite{Dan98}
the 3B effective potential in the hyperspherical coordinates is given as
a function of the hyperradius $\rho$
\be
V^{(R5)}(\rho) = \frac{-V_3 }{1+(\hat{\rho}/5)^3}~~.
\ee
In the second model (R2) of \cite{Tim00}
the potential is defined as
\be
V^{(R2)}(\rho)= - U_3 \exp(-\hat{\rho}^3)~~.
\ee
In these equations $\hat{\rho} = \rho/\rho_0$, where $\rho_0$= 5~fm and 1~fm
for R5 and R2 respectively.
The strength of the 3B effective potential is tuned to reproduce
the experimental three-body separation energy,
with V$_3$=--1.60 MeV and  U$_3$=--293.5 MeV.
The models R5 and R2 predict, with an $\alpha$ particle rms
matter radius of 1.49 fm, $^6$He rms matter radii of 2.50 fm
and 2.35 fm respectively.

\section{Results}

In this section we evaluate the elastic scattering differential
cross section for the scattering of protons on $^6$He at $E_{\rm
lab}=717$~MeV within the four-body FIA single scattering
approximation and Fixed scatterer approximation FSA.
In the case under study, the single scattering terms involves contributions
from the valence nucleons and from the  $^4$He core.

\subsection{p+cluster}

In the factorized impulse approximations discussed in the present
work,
the total T-matrix is expressed in terms of the on-shell matrix elements
of the two-body amplitudes evaluated at an appropriate momentum transfer
and energy. The $NN$ on-shell scattering amplitudes were obtained
from a realistic $NN$ Paris interaction, as in \cite{Cres99,Cres02,CresRNB}.

%%%%%%%%%%%%%%%%%%%%%%%%%%%%%%%%%%%%%%%%%%%%%%%%%%%%
%                 Cluster
\begin{figure}
{    \par\centering \resizebox*{0.9\columnwidth}{!}
     {\includegraphics{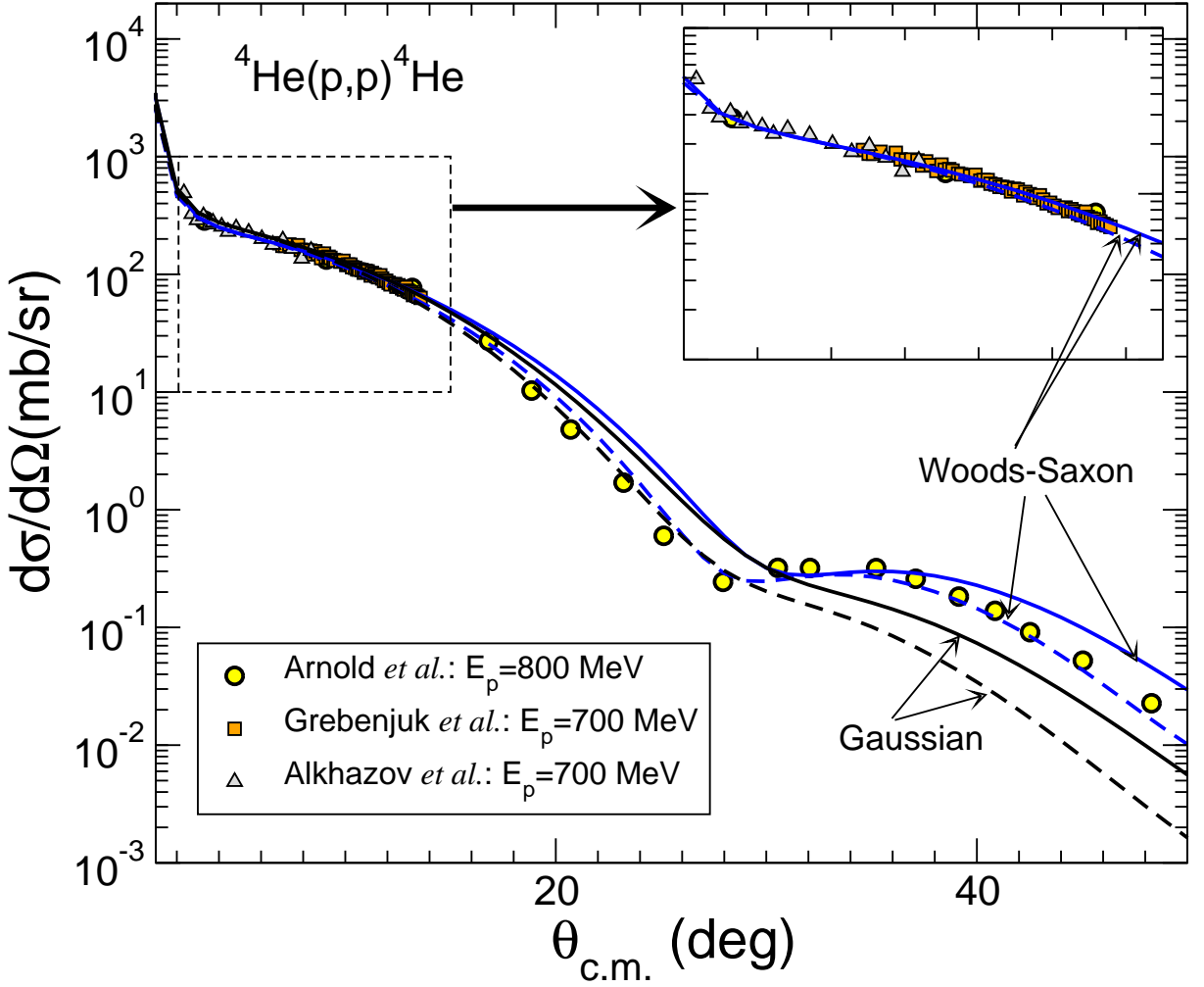}}                  \par}
\caption{\label{Fig:he4pp_e700}(Color online)
Calculated elastic scattering for $p+^4$He
at $E_p$=700 MeV (solid lines) and 800 MeV (dashed lines),
using a Gaussian and a Woods-Saxon parametrization
as described in the text. The data are taken from
Refs.~\cite{Alk97,Greb89,Arn79}}.
\end{figure}
%%%%%%%%%%%%%%%%%%%%%%%%%%%%%%%%%%%%%%%%%%%%%%%%%%%%%%%%%%%%%%%%%%

The two-body scattering amplitude for the scattering  $p+^4$He
was calculated with a phenomenological optical potential,
with parameters obtained by fitting
existing data for the elastic scattering
$p+^4$He  at  $E_{\rm p}=700$ MeV \cite{Alk97,Greb89}
and  800 MeV \cite{Arn79}.
We consider two different parametrizations for the $p+\alpha$
optical potential. In the first one,  based on the
work of Baldini-Neto {\em et al.} \cite{Bal03}
we take the optical potential as a sum
of two terms, real and imaginary, of Gaussian shape
\begin{equation}
U(r) = V_0 {\cal F}(r,r_0) + i W_0 {\cal F}(r,r_i)
\label{eq:potential}
\end{equation}
with ${\cal F}(r,r_x)=\exp(-r^2/r_x^2)$.  The values of the radii and the
real and imaginary depths were adjusted simultaneously in order to
minimize the $\chi^2$ with the experimental data. These fits were
performed with the computer code FRESCO \cite{Thom88}, version frxy.
The final values of these parameters where $V_0$=+73 MeV, $r_0$=1.37 fm,
$W_0$=-162 MeV and $r_i$=1.32 fm.

We have also taken a phenomenological potential of
Woods-Saxon (WS) shape, i.e.:
\be
U(r) = V_0 f(r,r_0,a_0) + i W_i f(r,r_i,a_i)
\ee
with $f(r,r_x,a_x) = [1 + \exp[(r - R_x)/a_x]^{-1}$,
where $R_x = r_x A^{1/3}$.

To reduce the number of degrees of freedom in the best fit search
we constrained the parameters to $r_0$=$r_i$ and $a_0$=$a_i$.
With this condition, a best-fit analysis of the data yields the values
$V_0$=+72 MeV,  $W_i$=-115 MeV,
$r_0$=$r_i$=1.14 fm, and  $a_0$=$a_i$=0.31 fm.
Note that with both geometries the real potential was positive in our
fits, indicating a dominance of the repulsive part in the $p+^{4}$He
interaction at these energies.
The data and calculated differential elastic cross sections
obtained with these parametrized  potentials for $p+^4$He at $E_p$=700
and 800 MeV are shown in Fig.~\ref{Fig:he4pp_e700}.
The solid and dashed lines,  corresponding to
$E_p$=700 MeV and 800 MeV respectively,
reproduce very well the forward scattering data
for both the Gaussian and Woods Saxon parametrization.
By contrast, the calculated differential cross section
tends to deviate from the data of \cite{Arn79} at larger angles
in the case of the Gaussian parametrization.
The WS optical model improves the fit in
this angular region.

\subsection{$p+^6$He}

The two-body $NN$ and $p+\alpha$ scattering amplitudes obtained by fitting
elastic data are now used to evaluate the total transition amplitude
for $p+^6$He scattering.

The Coulomb interaction was included in an approximate way as
summarized in appendix A.
The elastic scattering observables shown in here were evaluated using
relativistic kinematics as discussed in detail in appendix B. The relativistic
kinematic effects nevertheless were found to be small.
%------------------------------------------------------------------------
% Sensitivity with respect to the p + alpha OP
%------------------------------------------------------------------------
\begin{figure}
{    \par\centering \resizebox*{0.9\columnwidth}{!}
      {\includegraphics{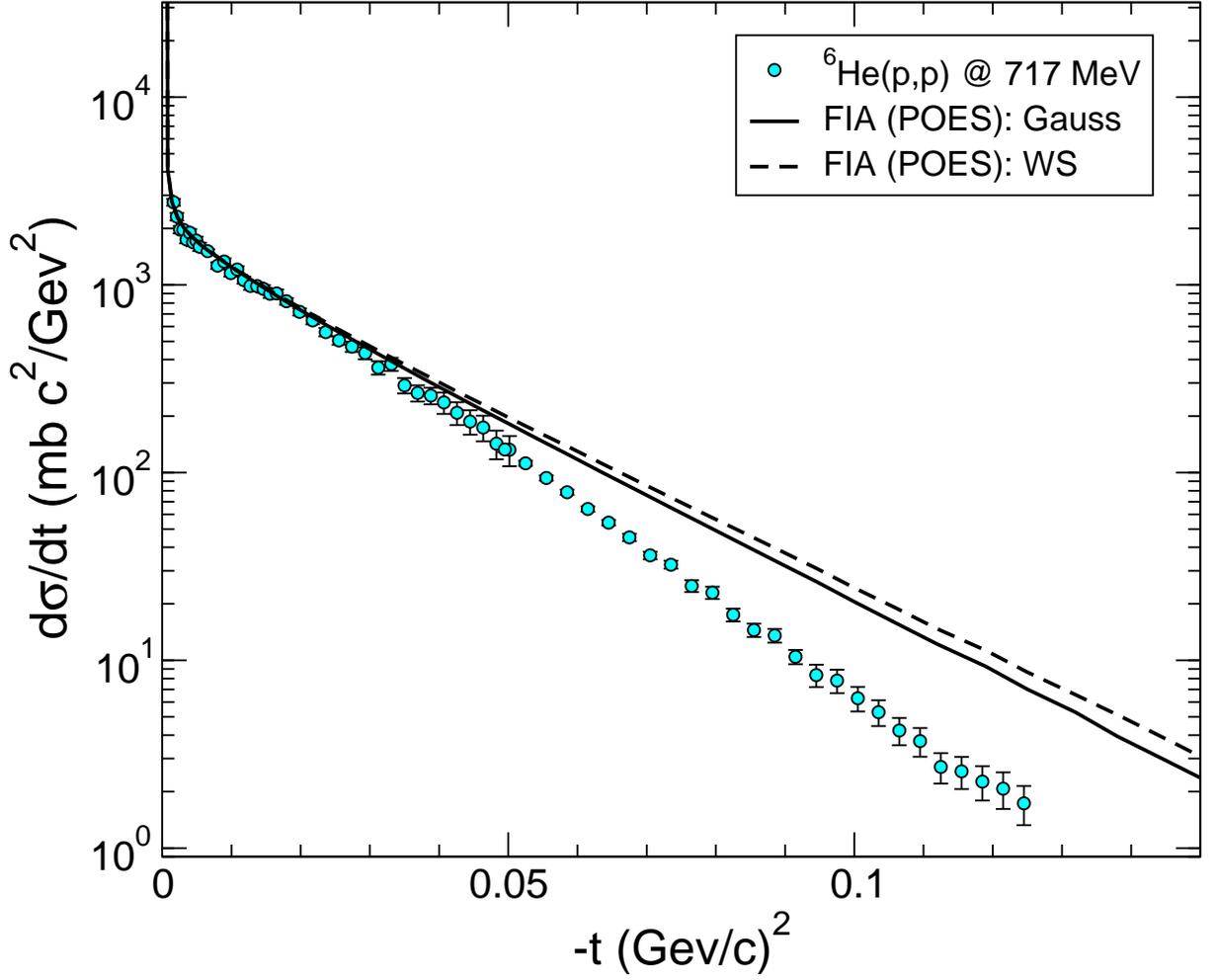}}         \par}
\caption{\label{Fig:he6-om}(Color online)
Calculated $p+^6$He elastic scattering at 717 MeV
for two different optical potentials for
$p+^4$He. The data is taken from \cite{Aks02, Ege02, Aks03}.
}
\end{figure}
%----------------------------------------------------------------
In Fig.~\ref{Fig:he6-om} we study
the dependence of the calculated differential cross section for
$p+^6$He elastic scattering  with respect to
the underlying $p+^4$He optical potential as a function of
the squared four momentum transfer $-t=\Delta^2$.
The solid and dashed lines
represent the  FIA-POES calculation using the Gaussian and WS Optical
model parametrizations for core scattering.
At higher momentum transfers the WS calculation gives a
slightly bigger reduction of the cross section.
The difference with respect to the Gaussian
parameterization is small, and both parametrizations predict
essentially the same differential cross section. The conclusions
of the present work are therefore essentially
independent of the underlying OM potential
for the scattering from the core. We shall be using for definiteness the
WS potential in all subsequent calculations.

%----------------------------------------------------------------
\begin{figure}
{   \par\centering \resizebox*{0.9\columnwidth}{!}
     {\includegraphics{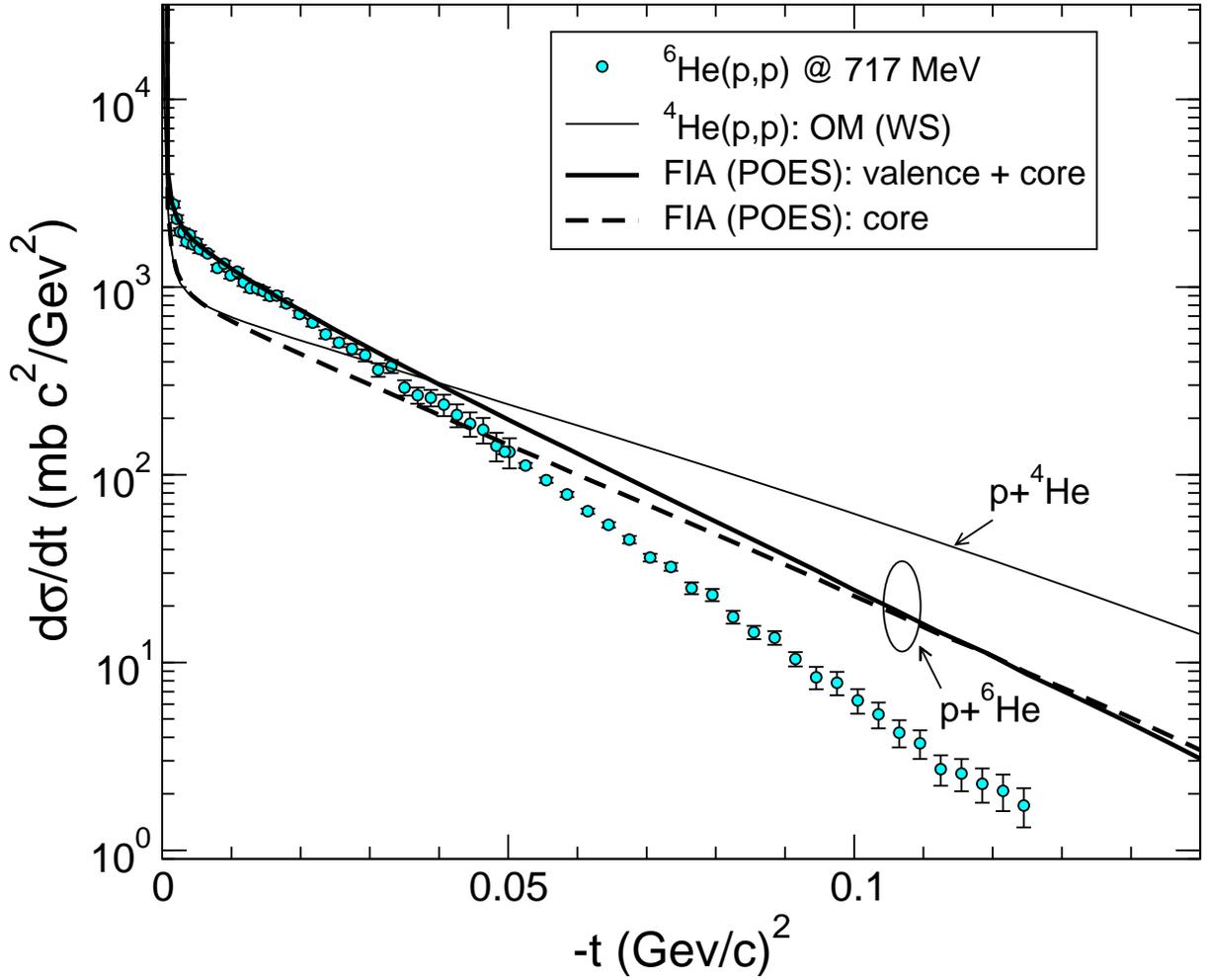}}       \par}
\caption{\label{Fig:he6pp_vst_recoil}(Color online)
Experimental and
calculated $p+^6$He elastic scattering at 717 MeV
 within the FIA-POES approximation. For comparison purposes,
the OM calculation for  $p+^4$He elastic scattering, at the
same energy per nucleon, is also included.}
\end{figure}
%----------------------------------------------------------------

In Fig.~\ref{Fig:he6pp_vst_recoil},
the thick solid line represents the calculated differential elastic cross
section for $p+^6$He using the FIA-POES approximation. The
dashed line was obtained  neglecting the single scattering contribution from
the valence neutrons. By comparing these two calculations,
one finds that the core contribution dominates
the large angle region but underpredicts the forward
differential cross section.
Also shown in the figure is the calculated elastic cross section
for $p+^4$He (thin solid line).
For values of $-t<0.03$~(GeV/c)$^2$ ($\theta_\mathrm{c.m.} < 11^\circ$)
the calculated $p+^6$He cross section is significantly
bigger than the $p+^4$He
cross section, providing a  good agreement with the
experimental data. Comparison
of the two FIA calculations indicate that this enhancement on the cross
section is mostly due to the presence of the valence neutrons, since
the FIA calculation with core contribution alone is very close to the
$p+^4$He distribution at these angles.

By contrast, for
larger values of momentum transfer, the few-body calculation is smaller
than the $p+^4$He curve.  One notes from the figure
that the two FIA calculations (core and core+valence single scattering)
are very close to each other at large momentum transfer, indicating that,
at least within the single scattering approximation, the depletion of
the cross section at larger angles is mostly a consequence of core recoil
effects.
Although the experimental $p+^6$He data  exhibit
a depletion of the cross section
with respect to the $p+^4$He distribution, the reduction predicted
by the FIA calculation is too small to explain the data.

%-----------------------------------------------------------------------
% Sensitivity to the IA
%------------------------------------------------------------------------
\begin{figure}
{      \par\centering \resizebox*{0.9\columnwidth}{!}
       {\includegraphics{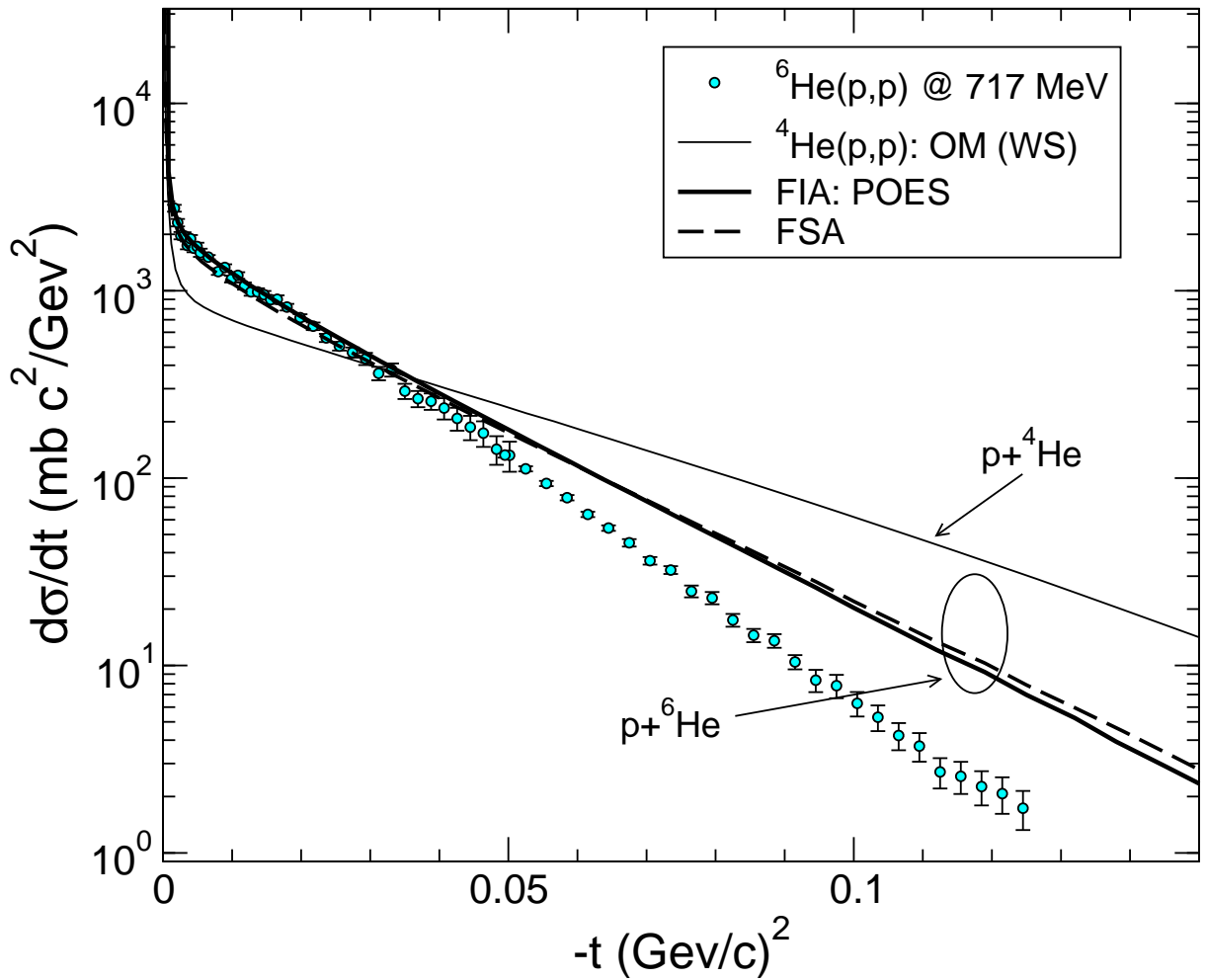}}   \par}
\caption{\label{Fig:he6pp_vst_models}(Color online)
Experimental and theoretical $p+^6$He elastic scattering at 717 MeV per
nucleon, for several scattering models discussed in the text. The calculated
$p+^4$He elastic distribution is included for comparison.}
\end{figure}
%----------------------------------------------------------------
In Fig.~\ref{Fig:he6pp_vst_models} we compare the calculated
$p+^6$He cross section using the FIA-POES (thick solid line),
the FIA-KL (dashed-dotted line) and the FSA (adiabatic)
approximation (long dashed).
The cross section calculated with the FIA-Rihan approximation is very
similar with the FIA-POES and therefore it is omitted from the figure.
The difference between the predicted differential cross
sections largely reflects the different rms matter radii.

The calculated differential cross section
using FIA (POES and Rihan) and FSA are very similar. The elastic
scattering observable calculated using the  KL impulse approximation
decays more slowly and the agreement with the data is comparatively even
poor at the large angle region.

%------------------------------------------------------------------------
% Sensitivity with respect to the 6He wavefunction
%------------------------------------------------------------------------
\begin{figure}
{   \par\centering \resizebox*{0.9\columnwidth}{!}
     {\includegraphics{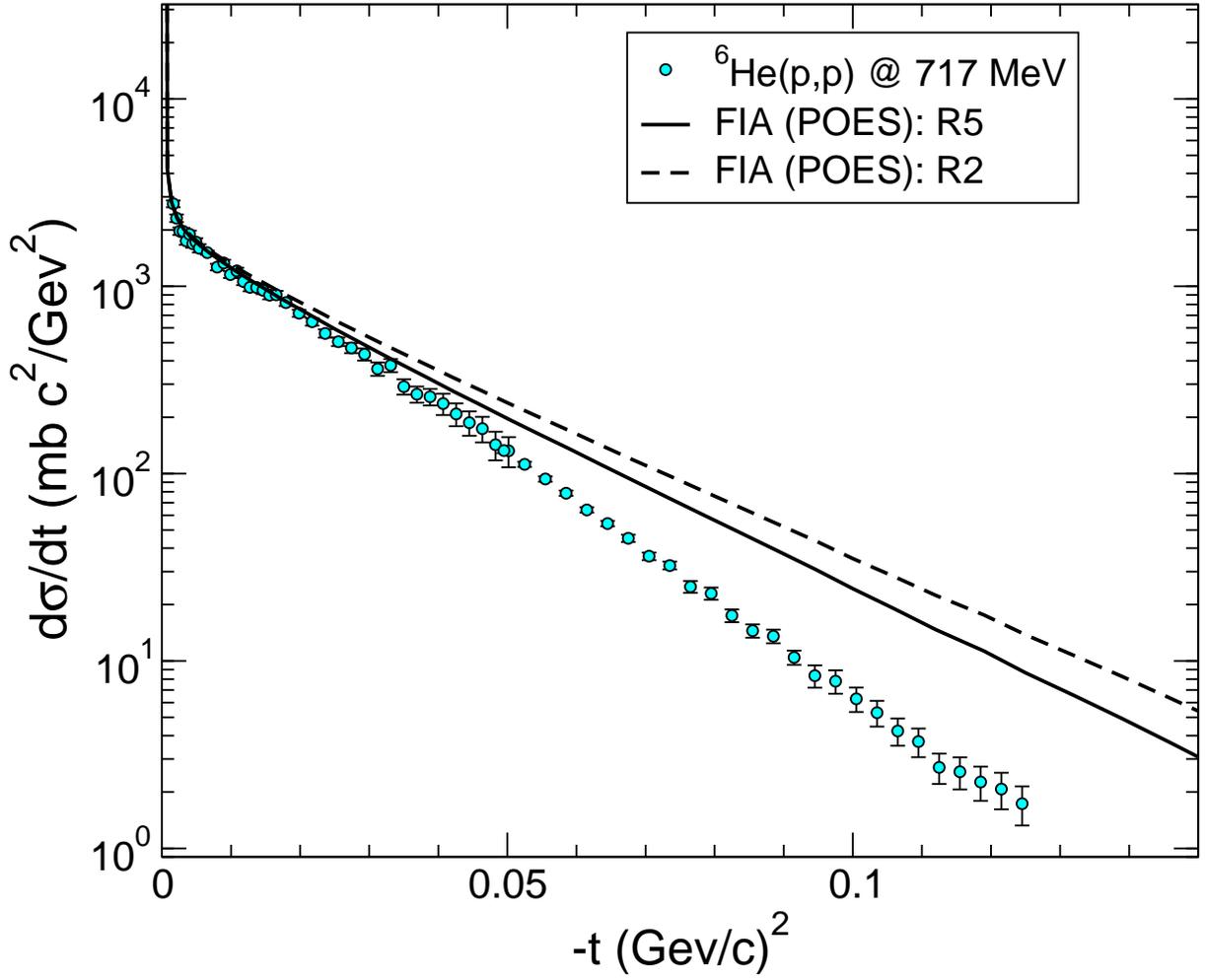}}      \par}
\caption{\label{Fig:he6pp_vst_struc}(Color online)
Experimental and calculated $p+^6$He elastic scattering at 717 MeV
for two different models for the $^6$He ground state.}
\end{figure}
%---------------------------------------------------------------
We now study the sensitivity of the results with respect to the $^6$He
wavefunction. In particular, we compare in Fig.~\ref{Fig:he6pp_vst_struc}
the FIA-POES calculations obtained with the R5 (thick solid line)
and R2 (dashed line)
$^6$He  models discussed in section \ref{sec6he}. Both models give
nearly identical results at low momentum transfer,
but the  R5, having a larger radius than R2,
exhibits a larger reduction  of the cross section at large scattering angles.
However, these differences do not appear to explain the disagreement
between the calculations and the data, and are much smaller than the
differences obtained when using different scattering approximations.

%------------------------------------------------------------------------
% Preliminary conclusions
%------------------------------------------------------------------------
 From the analysis presented in this section, we can conclude that the
factorized impulse approximations, in the different versions here presented,
describe fairly well the elastic scattering data for $p+^6$He at small
momentum transfer. In particular, they all show an increase of the cross
section with respect to the $p+^4$He scattering, showing the effect
of the valence neutrons  on the elastic scattering. At momentum transfers above
0.05~(GeV/c)$^2$ all the calculations tend to overestimate the data,
irrespective of the structure model used.

\section{Conclusions}
In this work we have reviewed and compared
several approaches for the scattering of a nucleus by a weakly bound composite,
based on the multiple scattering expansion of the total scattering amplitude,
namely, Factorized Impulse and Fixed Scatterer/Adiabatic Approximations.

The Factorized Impulse Approximation (FIA) neglects
the inter-cluster interaction  and approximates the
relative momentum between the loosely bound clusters.
As for the   Fixed Scatterer approximation (FSA) it
replaces the internal Hamiltonian by a constant.

As a common feature, all these
approaches express the single-scattering term as a product of a two-body
scattering amplitude for the scattering of the fragment times a structure
formfactor, which depends on the internal wavefunction of the composite system.
This factorized form provides a simple interpretation of the scattering
observable, by separating the role of the structure from the reaction dynamics.

These approaches have been then applied to the scattering of $^6$He on
protons at $E$=717 MeV per nucleon, using $p$-$n$ amplitudes derived
from a realistic $NN$ potential, and $p+^4$He amplitudes obtained
from an optical model fit of existing elastic data.
All the approximations here succeed in reproducing
the $ p+^6$He elastic data at small scattering angles. We found it crucial to
include the effect of the valence neutrons to the single scattering
contributions in order to explain the increase of
the cross sections with respect to the  $p+^4$He
elastic data at the same energy per nucleon.
At larger angles, both FSA and FIA calculations
tend to overestimate the data.

This disagreement  at larger angles
could be traced  to the neglect of higher order terms in
the T-matrix expansion and will be discussed elsewhere.

\begin{acknowledgments}
The financial support of Funda\c{c}\~{a}o para a Ci\^{e}ncia e a Tecnologia
from grant POCTI/FNU/43421/2001 and Ac\c c\~ao Integrada Luso-Espanhola
E-75/04 is  gratefully acknowledged.
A.M.M. acknowledges a postdoctoral grant by the Funda\c c\~ao para a Ciencia
e a Tecnologia (Portugal).
\end{acknowledgments}

\appendix

\section{Treatment of the Coulomb Potential}

The elastic differential angular distribution is evaluated from
the total scattering amplitude
\be
F^{\rm FIA} &=& {\cal N}_{12}^{1/2}
 \hat{f}_{2} (\omega_{12}, \vec{\kappa}_{12})
 \rho_{23,4}\left( \frac{m_3}{ M_{23} } \vec{\Delta},
\frac{m_4}{ M_{234} } \vec{\Delta} \right) \nonumber \\
 &+& {\cal N}_{13}^{1/2}
 \hat{f}_{3} (\omega_{13},   \vec{\kappa}_{13} )
\rho_{23,4}\left( \frac{m_2}{ M_{23} } \vec{\Delta},
\frac{m_4}{ M_{234} } \vec{\Delta} \right)
\nonumber \\
 &+& {\cal N}_{14}^{1/2}
 \hat{f}_{4} (\omega_{14},\vec{\kappa}_{14}    )
\rho_{23,4}\left( 0, \frac{M_{23}}{ M_{234} } \vec{\Delta} \right) ~~.
\nonumber \\
\label{Ffactorized4b}
\ee
In this equation, the scattering amplitude from the core includes the
Coulomb interaction.
The central scattering amplitude is taken as
\be
\hat{f}^c_{4} (\theta) &=& f^{pt}_C(\theta) \nonumber \\
&+& \frac{1}{Q_{14}}
\sum_L \exp(2 i \sigma_L) [ (L+1) T_4^{L+}(N)
\nonumber \\
&+&  ~~~~~~~~~~~  L T_4^{L-}(N) ] P_L(\cos\theta) ~~.
\label{fcoreCoulomb}
\ee
In Eq.~(\ref{fcoreCoulomb}), $f^{pt}_C(\theta)$  is the Coulomb scattering
amplitude due to a point charge $Z_4e$ and $Q_{14}$ is the asymptotic
projectile-core wave number. The $T_4^{L\pm}(N)$ are defined according to
\be
T_4^{L\pm}(N) = \frac{\exp\left[2i\delta_4^{L\pm}(N)\right] -1 }{2i} ~~,
\ee
where $L\pm$ denotes the orbital and total angular momenta,
$J=L\pm \frac{1}{2}$ and
$\delta_4^{L\pm}(N)$ are the Coulomb modified phase shifts
\cite{Crespo90}.
It follows from Eqs.~(\ref{Ffactorized4b}-\ref{fcoreCoulomb}) that the
point Coulomb scattering amplitudes appears multiplied by the structure
form factor
$\rho_{23,4}\left( 0, \frac{M_{23}}{ M_{234} } \vec{\Delta} \right)$.
Although this is an approximate treatment of the Coulomb interaction,
it will only have effects at very small projectile momentum transfer
and therefore does not modify the conclusions of the present work.

\section{Relativistic kinematic effects}

In this section we describe how the  expressions of sections III and IV
should be modified
in order to take into account relativistic kinematics. For simplification
we consider the 3-body case Eq.~(\ref{TFactorized3b}).
Expressions can be straightforward generalized to the 4-body case.
For simplicity,  we shall take in this section $c=1$.

Within MST, the projectile-target scattering scattering amplitude
is constructed from the projectile-subsystem scattering amplitude
as described in the text. We shall discuss  the relativistic kinematics
for the scattering of both the  composite and each subsystem.

Let us  consider the elastic scattering of projectile
labelled 1 with a  target $A$ with rest mass  $m_1$ and $m_A$ respectively,
\be
1 + A \rightarrow 1 + A
\ee
In our case $A$ represents the composite (2+3)
two-body system.
In relativistic kinematics one introduces
the Lorentz invariant Mandelstam variable $s_{1A}=(E_1+E_A)^2$  \cite{Joa87}.
The differential cross section for projectile-target with respect to
the four momentum transfer $t_{1A}$ is related to
the c.m. differential cross section as:
\be
\frac{d \sigma}{d t_{1A}} = \frac{ \pi}{k_1^2} \frac{d \sigma}{d \Omega}
~~,
\ee
where $t_{1A}=-|\Delta|^2 = 2 k_1^2 (\cos \theta_{\rm c.m.} - 1) $
and $k_1$ the projectile momentum in the projectile-target
CM frame, that can
be obtained from the Mandelstam variable  $s_{1A}$.
The cross section is evaluated from
the scattering amplitude,  related to the matrix elements of
the total transition amplitude according to \cite{Joa87}
\be
F &=& \frac{(2 \pi^2)^4}{\hbar v_1} k_1^2
\frac{dk_1}{d \omega} \langle  \vec{k}\,'_1 | T(\omega) |  \vec{k}_{1}\rangle
~~,
\label{Frelativistic}
\ee
with $\omega=\sqrt{s_{1A}} - m_1 - m_A$.

Let us now consider now the projectile scattering from subsystem ${\cal I}=2$.
As before, one introduces the Mandelstam variable for the interacting pair
$s_{12}$.
For simplification, we shall take the relative momentum of the interacting
pair $q_{23}$ nonrelativistically.
We shall consider in here the two situations where  $q_{23}=0$ (as in
KL and POES impulse approximation)
and $q_{23}\neq 0$ (as in  optimal approximation discussed by Rihan)
\begin{itemize}
\item
In the former case,  in the laboratory frame  $k_2 = k_3 = 0$ and
the Mandelstam variable $s_{12}$ can be readily evaluated
\be
s_{12} = {m}_{1}^2  + {m}_2^2
+ 2 {m}_2 ({m}_1  + T_1^{\rm Lab}) ~~.
\ee
\item
In the later case, where
$\vec{q}_{23}= \frac{1}{2} \frac{m_3}{M_{23}}\vec{\Delta}$,
it is more convenient to evaluate  $s_{12}$ in the
projectile-subsystem ${\cal I}=2$ CM frame. In this case,
\be
s_{12} = \left( \sqrt{m_1^2 + \vec{k}_1^2} +  \sqrt{m_2^2 + \vec{k}_2^2}
\right)^2 - ( \vec{k}_1 + \vec{k}_2 )^2
\ee
To evaluate $s_{12}$, one takes
$k_1$  from the Mandelstam invariant $s_{1A}$. The last term gives
\be
( \vec{k}_1 + \vec{k}_2 )^2 =
\frac{1}{2}\gamma_{12}^2 k_1^2 \left[ 1 + \cos\theta_{\rm c.m.} \right]
\ee
and
\be
k_2^2 = k_1^2 \left[ \chi_{12}^2 + \frac{1}{4} \gamma_{12}^2
- \gamma_{12} \chi_{12} \cos\theta_{\rm c.m.}   \right]
\ee
where $\chi_{12}=m_2/M_{23}  + \frac{1}{2}  \gamma_{12}$
\end{itemize}
The differential cross section for the scattering from subsystem
${\cal I}$ with respect to the four momentum transfer
$t_{12}$,  is related to the c.m. differential cross section as:
\be
\frac{d \sigma}{d t_{12}} = \frac{\pi}{{\cal Q}_{12}^2}
\frac{d \sigma}{d \Omega}_{12} ~~,
\ee
where $t_{12}=-|\kappa_{12}|^2 = 2 {\cal Q}_{12}^2(\cos \theta_{1,2} - 1)$
and ${\cal Q}_{12}$ the projectile momentum in the projectile-subsystem CM
frame that can be obtained from $s_{12}$.
The elastic scattering amplitude is related to the T-matrix elements through:
\be
f_{2} = \frac{(2 \pi^2)^4}{\hbar v_1} k_1^2
\frac{dk_1}{d \omega_{12} }
\langle {\cal \vec{Q}'}_{12} | \hat{t}_2 (\omega_{12})|
{\cal \vec{Q}}_{12}\rangle ~~,
\label{f2relativistic}
\ee
with $\omega_{12}=\sqrt{s_{12}} - m_1 -m_2 $ ~~.

\bibliographystyle{unsrt}
\bibliography{./mst}

\end{document}